# A general flux-Based Circuit Theory for Superconducting Josephson Junction Circuits

Yongliang Wang [a,b]*    *Member, IEEE*

[a] *Shanghai Institute of Microsystem and Information Technology (SIMIT), Chinese Academy of Sciences (CAS), Shanghai 200050, China*
[b] *CAS Center for Excellence in Superconducting Electronics, Shanghai 200050, China*

*Corresponding author. Tel.: +86 02162511070; Fax: +86 02162127493.

E-mail address: wangyl@mail.sim.ac.cn

ORCID: Yong-Liang Wang (0000-0001-7263-9493)

**Abstract**

   Superconducting quantum interference devices (SQUIDs), single flux-quantum (SFQ) logic circuits, and quantum Josephson junction circuits have been developed into a family of superconductor integrated circuit, and are widely applied for subtle magnetic-field measurements, energy-efficient computing, and quantum computing, respectively. They are Josephson junction networks composed of Josephson junctions and normal resistor-inductor-capacitor (RLC) components, working with the fluxoid-quantization principle and Josephson effects to achieve unique flux-modulated dynamics and characteristics; they react to the vector potential of magnetic fields rather than the electric potential. However, the conventional circuit diagrams and nodal analysis methods focus on the electric charges flowing though branches and nodes, ignoring dynamics of the magnetic fluxes flowing from loop to loop. This article introduces a general flux-based circuit theory to unify the analyses of Josephson junction circuits and normal RLC circuits. This theory presents a magnetic-flux-generator (MFG) concept to unify Josephson junctions and normal circuit elements, and abstract both Josephson junction circuits and normal RLC circuits as MFG network; it derives a general network equation to describe dynamics of Josephson junction circuits, and invents a kind of magnetic-flux flow (MFF) diagram to depict the working principles of magnetic-flux flows inside Josephson junction circuits. The flux-based theory is complementary to the conventional circuit theories in the design and analysis of superconductor integrated circuits.

# 1. Introduction

For nearly 60 years of development, various Josephson junction circuits have been developed for both analog and digital applications [1]. Superconducting quantum interference devices (SQUIDs) [2] are the typical Josephson junction circuits working as the ultra-sensitive flux-to-voltage converters; they are widely applied in subtle magnetic field measurement systems, such as magnetocardiogram and magnetoencephalogram [3], [4]. Single-flux-quantum (SFQ) circuits [5] are the large-scale digital Josephson junction networks, which are promising to develop ultra high-speed digital processors for power-efficient supercomputing systems [6], [7]. Furthermore, Josephson junction circuits in zero-voltage state can be used to design quantum electromagnetic circuits for quantum computing [8], [9]. Several typical SQUIDs [10-16] are illustrated in Fig. 1; they are simple networks of Josephson junctions connected in parallel or series by superconducting wires. SFQ logic circuits [17] are constructed with similar topology of SQUID circuits, as shown in Fig. 2. Furthermore, superconductor circuits will work with normal resistor-inductor-capacitor (RLC) elements in practical applications, as shown in Fig. 3. A rf-SQUID is coupled with a non-superconducting tank circuit in readout circuit, as shown in Fig. 3(a); dc-SQUID and bi-SQUID will be connected through normal conductor wires to the room-temperature semiconductor amplifiers, as depicted in Fig. 3(b) and (c); the dro-SQUID shown in Fig 3(d) contains both superconducting and normal loops.

Superconductor wires and Josephson junctions are the two kinds of superconductive elements brought by Josephson junction circuits [18]. Superconductor wires are magnetic-field couplers, inside which cooper pairs in condensed state are described with one macroscopic wave-function $\psi$, as shown in Fig. 4(a), and the phase $\varphi$ varies with the vector potential $A$ of magnetic field [19]. A typical superconductor-insulator-superconductor (SIS) Josephson junction is illustrated in Fig. 4(b), where there is a Josephson current $i_J$ tunneling through the insulator [20], $i_J = I_0\sin\theta$; $\theta$ is the phase difference and $u$ is the voltage difference between two superconductors, $d\theta/dt = 2\pi u/\Phi_0$. In practical applications, the Josephson junction is usually shunted by a resistor and is called the resistively-capacitively-shunted junction (RCSJ) [21].

Based on the phase-voltage relation in Josephson equations [19], it is not difficult to use nodal phases or the equivalent variables to redefine nodal voltages and transform the conventional modified-nodal analysis (MNA) method [22] into the so-called phase-based MNA methods or the equivalents [23], [24]. By using the phase-based MNA methods, the simulation tools for Josephson junction circuits, such as, PSCAN2[23], JSIM [25], JSPICE [26], JoSIM [27], and APLAC [28], are developed on the basis of the conventional SPICE (simulation program with integrated circuit emphasis) tool. Given the circuit diagram of a Josephson junction circuit, one can use the phase-base method to set up the circuit equations with the netlist extracted from the circuit diagram and calculate the real-time responses of all the circuit elements by finding the numerical solutions; those jobs are automatically done by SPICE tools.

However, we find that conventional circuit diagrams are inconvenient for modeling the Josephson junction circuits with various flux inputs, and the working principle of

Josephson junction circuits is not well interpreted with all the solutions of nodal phases found by MNA methods; lots of questions are encountered in the analyses of Josephson junction circuits. For example, node phases in voltage state will keep increasing to infinity; what is physical meaning of the nodal phases for circuit elements; the uniqueness of nodal phases will be broken, if a loop is trapped with flux quanta; how to describe the trapped flux quanta with circuit elements and make the circuit diagram comply with the fluxoid-quantization law (FQL) ? why can't we directly tell the '0' and '1' states of SFQ logics [29] with nodal phases; what is the real entity of SFQ logic bits; how does it preserve the logic state after Josephson junctions return to zero-voltage state? The reason may be that those phase-based MNA methods and simulation tools provide merely the analyses in mathematics rather than physics for Josephson junction circuits.

This article introduces a general flux-based circuit theory to unify Josephson junction circuits and normal RLC circuits. In this theory, a magnetic-flux-generator (MFG) concept [30] is presented to unify the definitions of Josephson junctions and normal circuit components; a general MFG network abstraction is applied to unify Josephson junction circuits and normal RLC circuits; a general network equation is derived to describe the dynamics of both superconducting and normal RLC circuits, and a kind of magnetic-flux-flow (MFF) diagram [31] is invented to depicts working principles of magnetic-flux flows inside Josephson junction circuits.

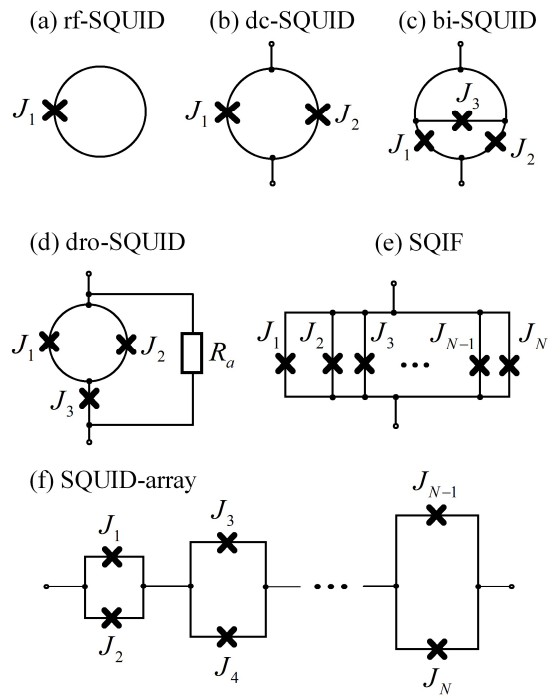

Fig. 1. Typical SQUIDs: (a) radio-frequency (rf) SQUID [10]; (b) direct-current (dc) SQUID [11],[12]; (c) bi-SQUID [13]; (d) double-relaxation-oscillation (dro) SQUID [14]; (e) superconducting quantum interference filter (SQIF) [15]; (f) dc-SQUID array [16]. A cross "×" in diagrams represents a resistively shunted Josephson junction

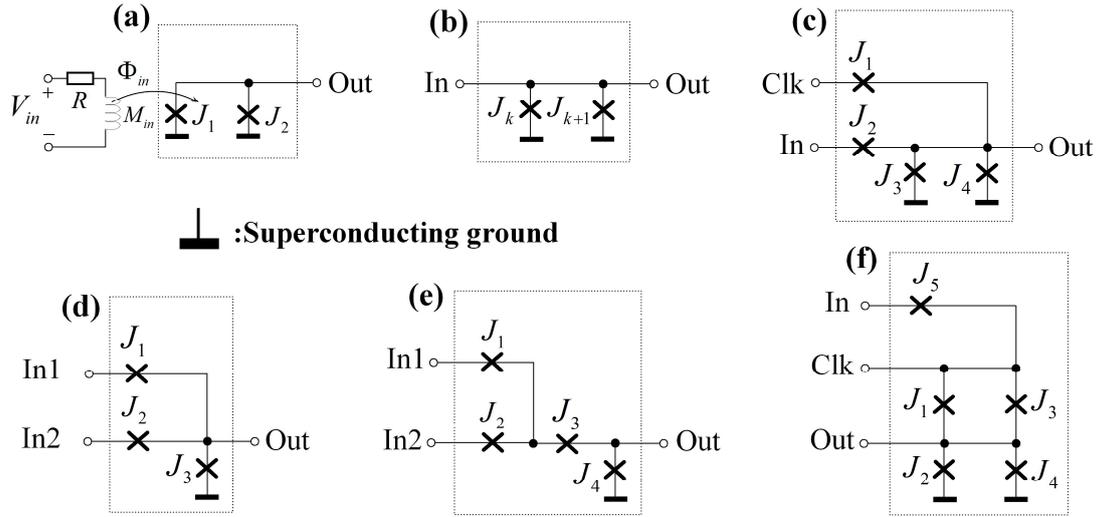

Fig. 2. Typical SFQ digital circuits [17]: (a) TTL-to-SFQ convertor; (b) Josephson transmission line (JTL); (c) Josephson D-type flip-flop (DFF); (d) Josephson AND-gate, (e) Josephson OR-gate, (f) Josephson NOT-gate.

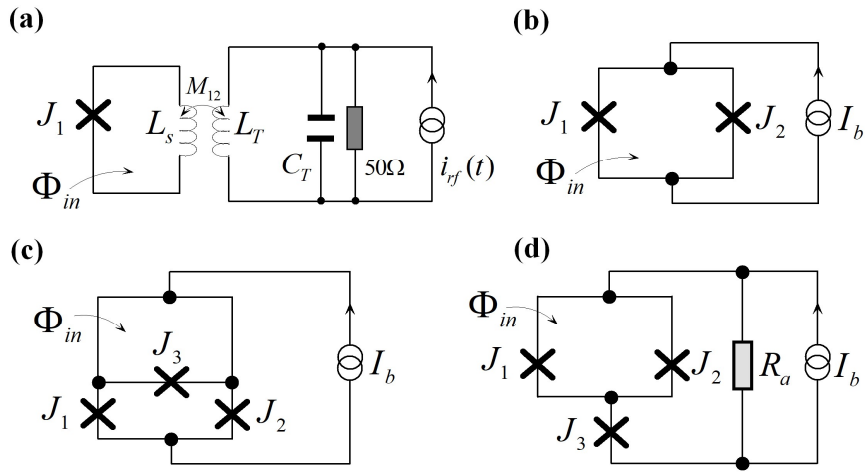

Fig. 3. SQUID bias circuits [18]: (a) rf-SQUID circuit tuned by a radio-frequency current $i_{rf}(t)$; (b) dc-SQUID circuit biased by a dc current; (c) current-biased bi-SQUID circuit; (d) current-biased dro-SQUID circuit.

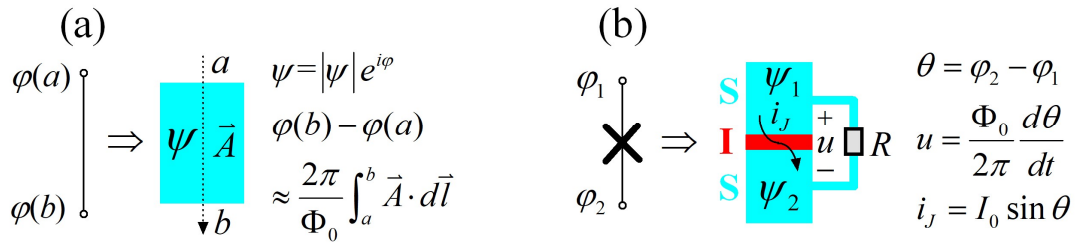

Fig. 4. Superconductive elements [19]: (a) superconductor wire; (b) Josephson junction.

## 2. Theory

*2.1 Charge-flux duality*

Due to the charge-flux duality in one circuit [32], an electric circuit is both an electric-charge distribution system and a magnetic-flux distribution system; it viewed from electric field is a network where electric charges are transferred from node to node, and is also a network where magnetic fluxes are transferred from loop to loop, viewed from magnetic field [31], as illustrated in Fig. 5. Different views require different circuit diagrams and different analysis methods, correspondingly.

Conventional circuit diagrams are drawn by connecting branches to nodes; they are the roadmap of charges flowing among nodes. The MNA method focuses on nodes and branches, and will divide the magnetic flux in loops into parts and hide them in branches with inductors [22-24] to analyze the distribution of electric charges. Accordingly, conventional circuit diagrams and the MNA method are cooperated to study the electric-charge distribution dynamics of the circuit; they are not suitable for Josephson junction circuits that work as magnetic-flux distribution networks where superconducting currents react to the magnetic fields and preserve flux quanta with superconducting loops.

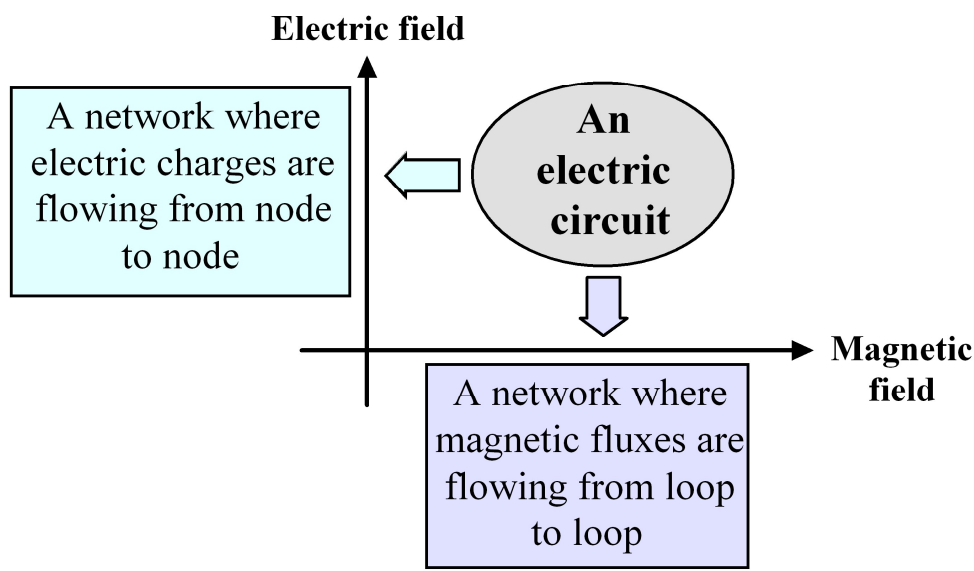

Fig. 5. Charge-flux duality of electric circuits: an electric circuit is an electric-charge distribution network where electric charges are flowing from node to node, and is meanwhile a magnetic-flux distribution network where magnetic fluxes are flowing from loop to loop.

*2.2 Overview of flux-based circuit theory*

In our general flux-based theory, the major concepts and models to unify Josephson junction circuits with normal RLC circuits are illustrated in Fig. 6. The key approach of this method is to abstract both superconducting and normal RLC circuits as MFG networks. A MFG network consists of only two kinds of components: the one is MFG which is the common model of current-biased Josephson junctions and current-biased resistor-capacitor (RC) components; another one is independent loop working as the magnetic-flux-container that stores the outputs of MFGs. Finally, Josephson junction circuits and normal RLC circuits are uniformly abstracted as MFG networks; their dynamics are described by the general network equation, and graphically depicted with MFF diagrams.

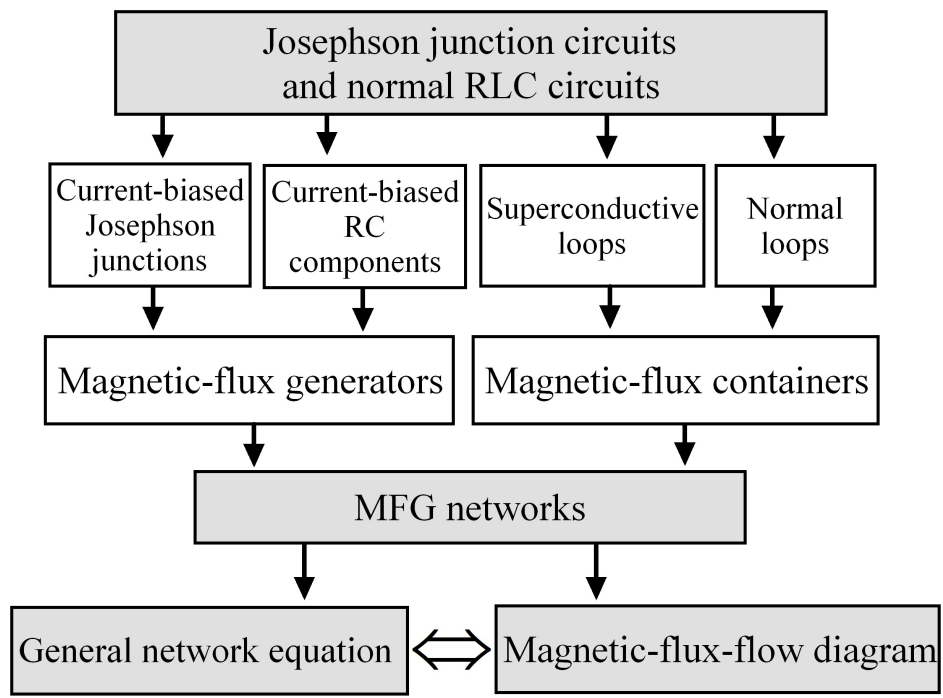

Fig. 6. Procedure of deriving the general network equation to unify Josephson junction circuits and normal RLC circuits.

*2.3 Magnetic-flux generators*

In Josephson junction circuits and normal RLC circuits, resistors, capacitors, Josephson junctions, and current sources are inserted in the breakpoints of normal or superconductive wires (inductors are treated as wires); they will create the phase and potential differences between two wires. Three kinds of breakpoints are illustrated in Fig. 7.

1) A breakpoint of normal wires is illustrated in Fig. 7(a), where the combination of a resistor, a capacitor, and a bias current in parallel is called the current-biased RC component. The RC component will create a potential difference $u_{EL}$ between two terminals when it driven by the bias current source $i_b$ and the branch current $i_{EL}$; its equivalent circuit is shown in Fig. 7(b), where, $i_C$ is the displacement current bypassed by the capacitance, and $i_R$ is the normal current through the resistor.

2) A breakpoint of superconductor wires is shown in Fig. 7(c), where a current-biased RC component is inserted, and creates a quantum-phase difference $\theta_{EL}$ between two superconductors. The equivalent is shown in Fig. 7(d), which is same with the one shown in Fig. 6(b).

3) A breakpoint inserted with a current-biased Josephson junction is depicted in Fig. 7(e). The RCSJ model of Josephson junction is shown in Fig. 7(f), which is equivalent to a RC component shunted by a Josephson current.

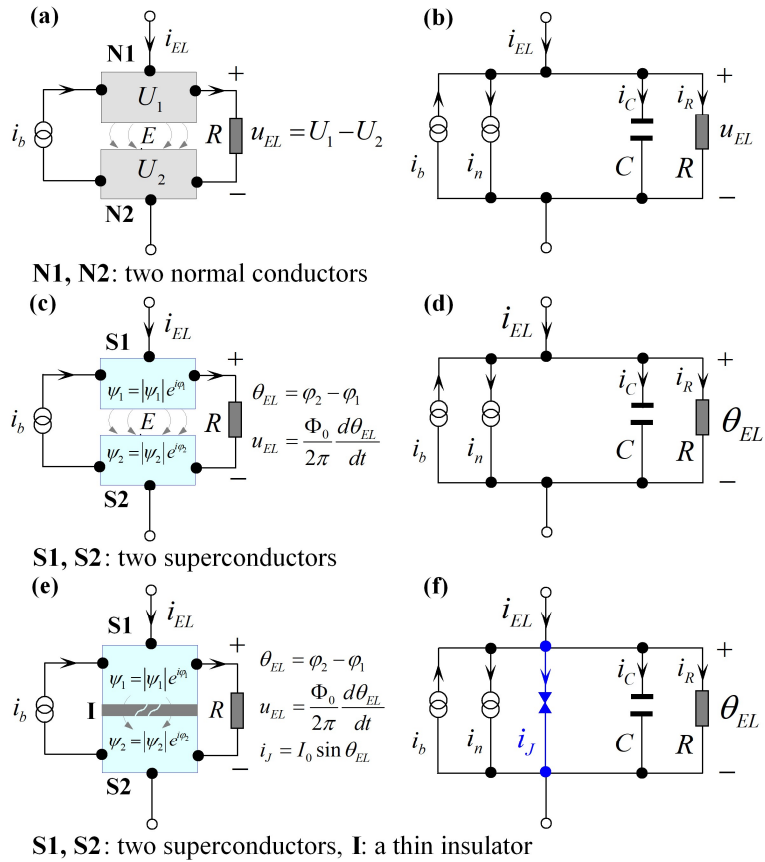

Fig. 7. Three types of components in the breakpoints of wires: (a) current-biased RC component inserted between two normal conductors, where $U_1$ and $U_2$ are the electric potentials; $i_b$ is the external bias current source; $R$ is the shunt resistor; its (b) equivalent circuit is a two-terminal network with a resistor and a capacitor connected in parallel, and $i_n$ is the current noise. (c) Current-biased RC component inserted between two superconductors and (d) the equivalent circuit, where $\varphi_1$ and $\varphi_2$ are the macroscopic quantum phases. (e) Current-biased Josephson junction and the (f) equivalent circuit which is a normal RC component shunted by a Josephson current $i_J$.

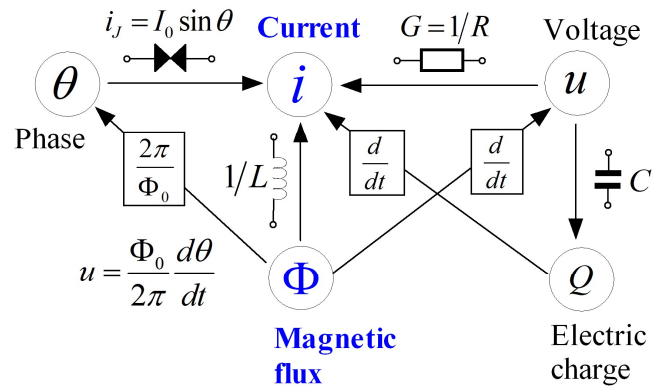

Fig. 8. Variables for defining superconductive and normal circuit elements.

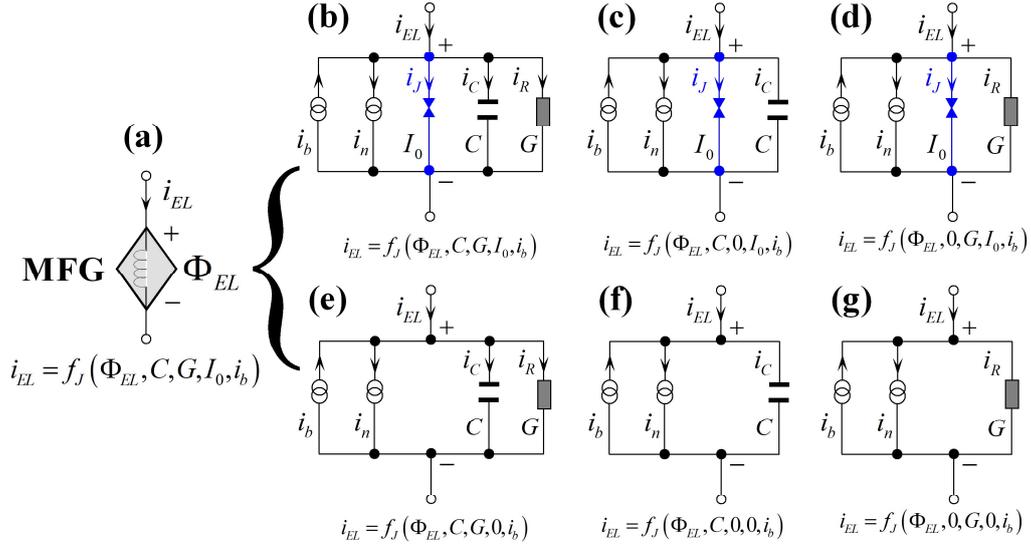

Fig. 9. Electric symbol of MFGs and the equivalent circuits: (a) two-terminal electric symbol of MFGs; (b) current-biased RCSJ, where $G$ is the conductance of $R$, $G = 1/R$; (c) current-biased capacitively-shunted junction (CSJ); (d) current-biased resistively-shunted junction (RSJ); (e) current-biased RC component; (f) current-biased capacitor; (g) current-biased resistor.

The elements and variables for modeling three types of RC components are exhibited in Fig. 8. We can use current ($i$) and flux ($\Phi$) as the common variables to define all the elements. Therefore, we introduce a nominal flux [32] $\Phi_{EL}$ to redefine the $u_{EL}$ and phase $\theta_{EL}$ at two terminals, and

$$\theta_{EL} = 2\pi \frac{\Phi_{EL}}{\Phi_0}; u_{EL} = \frac{d\Phi_{EL}}{dt} \tag{1}$$

where $\Phi_0 = 2.07 \times 10^{-15}$ Wb.

With $\Phi_{EL}$ as the output, all the RC components work as the magnetic-flux-generators (MFGs), which implement current-to-flux conversion in both superconducting and normal circuits. The symbol of MFGs and typical equivalent circuits of MFGs are exhibited in Fig. 9. Different MFGs share a general current-flux function [31] as

$$\begin{aligned} i_{EL} &= f_J(\Phi_{EL}, C, G, I_0, i_b) \\ &\triangleq C\frac{d^2\Phi_{EL}}{dt^2} + G\frac{d\Phi_{EL}}{dt} + I_0 \sin\left(2\pi \frac{\Phi_{EL}}{\Phi_0}\right) + i_n - i_b \end{aligned} \tag{2}$$

The configuration of $I_0$, $G$, $C$ tells the equivalent circuit of MFGs. For instance, a MFG with a none-zero $I_0$ is a Josephson junction; otherwise, it is a normal RC component.

The current-flux function of MFGs can be further depicted with transfer function diagrams shown in Fig. 10. A MFG with $C \neq 0$ is a second-order system, and the one with $C = 0$ is a first-order system, where, the Josephson current is an energy storage that creates a cosine-shape potential [19] for the RC component.

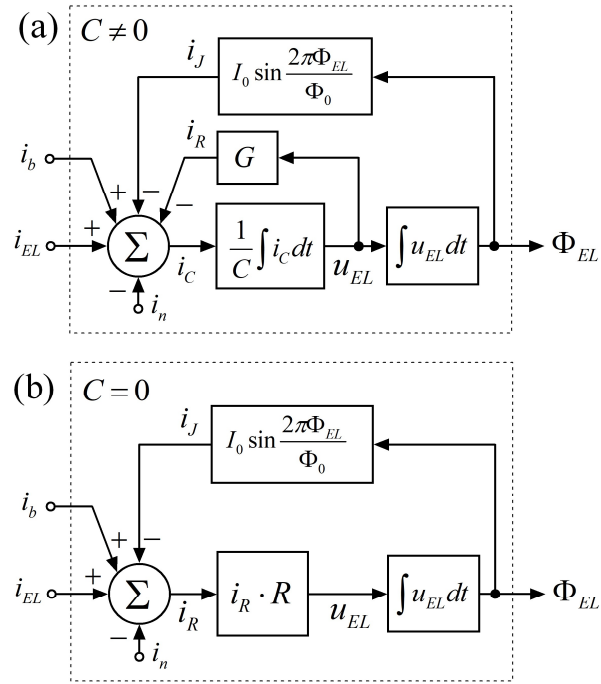

Fig. 10. (a) Transfer function diagrams of an MFG with $C \neq 0$, and (b) the one with $C = 0$.

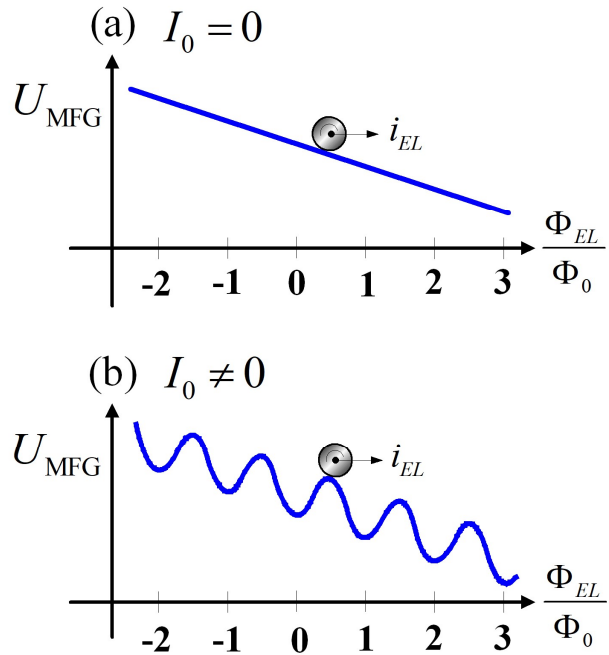

Fig. 11. Mechanical analog of an MFG: (a) a ball rolling in a potential set by $i_b$, and the one rolling in a wash-board style potential set by $i_b$ and $i_J$.

The dynamic of MFGs can be illustrated with the mechanical analog shown in Fig. 11. A MFG is analogous to a particle which mass is $C$, and the factor of moving resistance is $G$; it is dragged by a force $i_{EL}$, and rolling in a potential $U_{MFG}$. The $U_{MFG}$

is created by currents $i_J$ and $i_b$, and

$$U_{MFG} = \int_0^{\Phi_{EL}} (I_0 \sin \frac{2\pi \Phi_{EL}}{\Phi_0} - i_b) d\Phi_{EL}$$
$$= \frac{\Phi_0 I_0}{2\pi} (1 - \cos \frac{2\pi \Phi_{EL}}{\Phi_0}) - i_b \Phi_{EL}$$
(3)

A normal MFG with a $I_0=0$ is equivalent to the rigid ball rolling in a slope shown in Fig. 11(a), while a MFG with a $I_0 \neq 0$ is analogous to a ball rolling in a wash-board style potential [19], as shown in Fig. 11(b). The ball in Fig. 11(a) will be steady at any place, when it is stopped, which exhibits the behaviors of normal MFGs in the non-superconducting circuits. The one in Fig. 11(b) will be steady at the bottom of the potential valleys which are periodically located with an interval of $1\Phi_0$. This case simulates the behaviors of Josephson junctions in SFQ circuits. In SQUID circuits, Josephson junctions are in voltage state, and will keep rolling in the wash-board potential; the average speed of the ball is exactly the average voltage of the Josephson junction. In a superconducting qubit circuit, the Josephson junction in zero-voltage state is a ball swinging in the potential well created by Josephson current [33].

*2.4 Magnetic-flux containers*

A closed loop consists of $N$ MFGs is connected by $N$ pieces of wires, as demonstrated in Fig. 12(a). It is a superconducting loop, if wires are superconductors and all the MFGs are Josephson junctions; therefore, the macroscopic wave functions inside wires are coherent and unique [19], and the total variation of phases along the loop must be $2n\pi$ ($n$ is an integer); the FQL is expressed as follows:

$$\sum_{i=1}^{N}(\varphi_i(a_i) - \varphi_i(b_i)) + \sum_{i=1}^{N} \sigma_i \theta_{ELi} = 2n\pi$$
$$\Rightarrow \Phi_m + \sum_{i=1}^{N} \sigma_i \Phi_{ELi} = n\Phi_0; n \in \mathbb{Z}$$
(4)
$$\sigma_i = \begin{cases} 1; & \text{if } i_m \text{ enters '+' of MFG-}i \\ -1; & \text{if } i_m \text{ enters '−' of MFG-}i \end{cases}$$

where $\Phi_m$ is the total flux coupled by the loop.

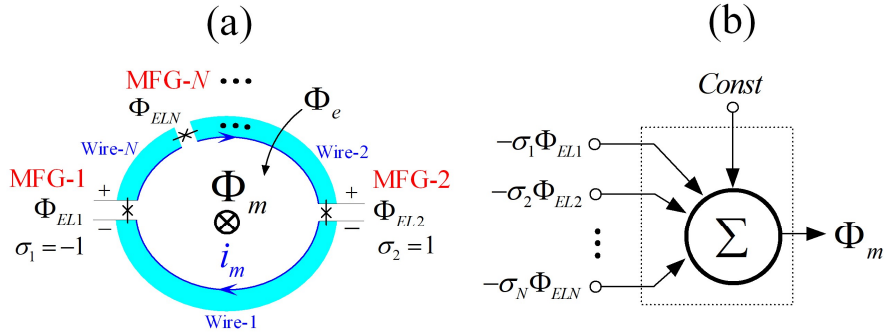

Fig. 12. (a) A loop consisting of $N$ MFGs, where a loop-current $i_m$ is circulating clockwise along the loop; (b) transfer function of the loops.

This loop becomes a non-superconducting loop, if any one of MFGs is a normal RC component; according to the Faraday's law of electromagnetic induction, we can have

$$\frac{d\Phi_m}{dt} + \sum_{i=1}^{N}\sigma_i u_{ELi} = 0$$
$$\Rightarrow \Phi_m + \sum_{i=1}^{N}\sigma_i \Phi_{ELi} = r\Phi_0; r \in \mathbb{R} \quad (5)$$

The FQL is exactly the Faraday's law of electromagnetic induction when the constant is specified as $n\Phi_0$. Thus, the flux-conservation law (FCL) [31] that is suitable for both superconducting and normal loops is rewritten as

$$Const - \sum_{i=1}^{N}\sigma_i \Phi_{ELi} = \Phi_m$$
$$Const = \begin{cases} n\Phi_0 (n \in \mathbb{Z}); \text{ for superconductive loops} \\ r\Phi_0 (r \in \mathbb{R}); \text{ for normal loops} \end{cases} \quad (6)$$

In this FCL, all the variables achieve practical physical meanings as follows:
1) the *Const* is the initial flux trapped in the loop;
2) the $\Phi_{ELi}$ records throughput of MFG-*i* to the loop, and the $\sigma_i$ indicates the direction of the flow;
3) the $\Phi_m$ is the total magnetic flux stored in the loop.

The loop shown in Fig. 12(a) looks like a magnetic-flux container; it collects all the flux inputs from MFGs and preserves them as $\Phi_m$, as illustrated in Fig. 12(b). To count the number of flux quanta contained in the loop, the $\Phi_m$ is normalized as $\chi$,

$$\chi \equiv \frac{\Phi_m}{\Phi_0} \quad (7)$$

The $\chi$ will be approximate to an integer in a superconducting loop, if all the MFGs are steady at the bottom of the potential, as shown in Fig. 11(b). In SFQ circuits, the '0' and '1' values of $\chi$ directly define the '0' and '1' states of SFQ logics.

For a standalone loop with a self-inductance $L_m$, the $\Phi_m$ is preserved by its loop current $i_m$ as

$$\Phi_m = L_m i_m - \Phi_e \quad (8)$$

where $\Phi_e$ is the total external flux applied to the loop, which is canceled by the self-induced flux.

*2.5 General equation of MFG networks*

A MFG network contains a group of independent loops [34], [35] (at least one loop), where MFGs are inserted in branches. Loops are either directly coupled by sharing branches or magnetically coupled through mutual inductances. If there is a given MFG network containing $P$ independent loops and $Q$ MFGs, we can depict the components and the topology information of the MFG network as shown in Fig. 13. In Fig. 13(a), each loop, Loop-*i* ($1 \leq i \leq P$), contains a total coupled flux $\Phi_{mi}$ and a loop-current $i_{mi}$ corresponding to an external flux input $\Phi_{ei}$; in Fig. 13(b), each MFG, MFG-*j* ($1 \leq j \leq Q$) outputs flux $\Phi_{ELj}$ corresponding to the element current $i_{ELj}$.

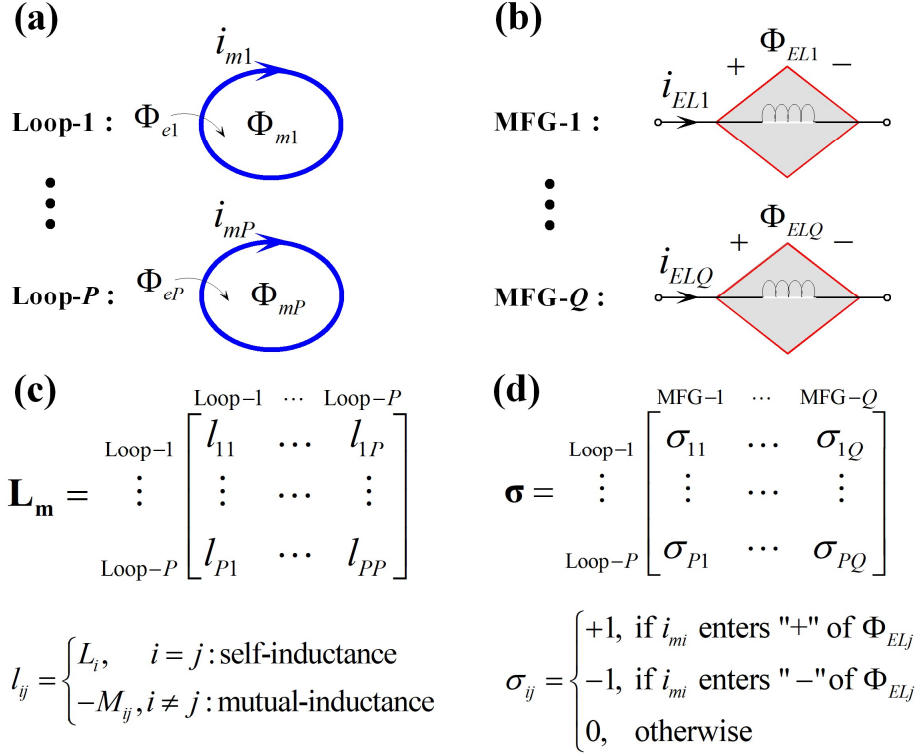

Fig. 13. Loops and MFGs in a given Josephson junction circuit: (a) independent loops; (b) MFGs; (c) Loop-to-loop inductance matrix $\mathbf{L_m}$; (d) Loop-to-MFG incidence matrix $\boldsymbol{\sigma}$.

The magnetic couplings of loops are defined by a loop-to-loop inductance matrix $\mathbf{L_m}$, as illustrated in Fig. 13(c), where the $L_i$ is the self-inductance of Loop-$i$, and $M_{ij}$ is the mutual-inductance between Loop-$i$ and Loop-$j$ ($1 \leq i, j \leq P$); the minus sign before $M_{ij}$ indicates that the flux contributed by the mutual-inductance is opposite to the one by the self-inductance. Meanwhile, the electric couplings between loops and MFGs are described by a loop-to-MFG incidence matrix $\boldsymbol{\sigma}$, as illustrated in Fig. 13(d).

The FCL in (6) for $P$ loops is expressed in matrix as

$$\mathbf{Const} - \boldsymbol{\sigma}\boldsymbol{\Phi}_{\mathbf{EL}} = \boldsymbol{\Phi}_{\mathbf{m}}$$
$$\begin{cases} \boldsymbol{\Phi}_{\mathbf{EL}} = \begin{bmatrix} \Phi_{EL1} & \cdots & \Phi_{ELQ} \end{bmatrix}^T \\ \boldsymbol{\Phi}_{\mathbf{m}} = \begin{bmatrix} \Phi_{m1} & \cdots & \Phi_{mP} \end{bmatrix}^T \\ \mathbf{Const} = \begin{bmatrix} Const_1 & \cdots & Const_P \end{bmatrix}^T \end{cases} \qquad (9)$$

and the expression in (8) for $P$ loops is written in matrix as

$$\boldsymbol{\Phi}_{\mathbf{m}} = \mathbf{L_m}\mathbf{i_m} - \boldsymbol{\Phi}_{\mathbf{e}}$$
$$\begin{cases} \mathbf{i_m} = \begin{bmatrix} i_{m1} & \cdots & i_{mP} \end{bmatrix}^T \\ \boldsymbol{\Phi}_{\mathbf{e}} = \begin{bmatrix} \Phi_{e1} & \cdots & \Phi_{eP} \end{bmatrix}^T \end{cases} \qquad (10)$$

According to the Kirchhoff's current law, the relation between branch currents and loop currents is

$$\mathbf{i_{EL}} = \boldsymbol{\sigma}^T \mathbf{i_m} \qquad (11)$$

Finally, we can derive the flux-current function of loops from (9)-(11) as

$$\mathbf{i_{EL}} = \boldsymbol{\sigma}^T \mathbf{L_m^{-1}}(\mathbf{Const} + \boldsymbol{\Phi_e}) - \boldsymbol{\sigma}^T \mathbf{L_m^{-1}} \boldsymbol{\sigma} \boldsymbol{\Phi_{EL}} \qquad (12)$$

where, $\boldsymbol{\sigma}^T$ is the transpose of $\boldsymbol{\sigma}$, and $\mathbf{L^{-1}_m}$ is the inverse matrix of $\mathbf{L_m}$.

Referring to the current-flux function of MFGs, we can derive the general network equation of MFG networks as follow:

$$\mathbf{C}\frac{d^2\boldsymbol{\Phi_{EL}}}{dt^2} + \mathbf{G}\frac{d\boldsymbol{\Phi_{EL}}}{dt} + \mathbf{I_0}\sin\left(\frac{2\pi\boldsymbol{\Phi_{EL}}}{\Phi_0}\right) + \boldsymbol{\sigma}^T\mathbf{L_m^{-1}}\boldsymbol{\sigma}\boldsymbol{\Phi_{EL}} \\ = \mathbf{i_b} - \mathbf{i_n} + \boldsymbol{\sigma}^T\mathbf{L_m^{-1}}(\mathbf{Const} + \boldsymbol{\Phi_e}) \qquad (13)$$

where, $\mathbf{i_b}$, $\mathbf{i_n}$ are the current vectors, $\mathbf{C}$, $\mathbf{G}$, and $\mathbf{I_0}$ are the parameter matrices of MFGs,

$$\begin{cases} \mathbf{i_b} = \begin{bmatrix} i_{b1} & \cdots & i_{bQ} \end{bmatrix}^T \\ \mathbf{i_n} = \begin{bmatrix} i_{n1} & \cdots & i_{nQ} \end{bmatrix}^T \\ \mathbf{C} = diag\{C_1, \cdots, C_Q\} \\ \mathbf{G} = diag\{G_1, \cdots, G_Q\} \\ \mathbf{I_0} = diag\{I_{01}, \cdots, I_{0Q}\} \end{cases} \qquad (14)$$

This network equation exhibits that:
1) the **Const** is merely an offset to $\boldsymbol{\Phi_e}$; **Const** and $\boldsymbol{\Phi_e}$ are turned into an offset to $\mathbf{i_b}$;
2) the item $\boldsymbol{\sigma}^T\mathbf{L^{-1}_m}\boldsymbol{\sigma}\boldsymbol{\Phi_{EL}}$ exhibits the interferences of MFGs;
3) the nonzero $\mathbf{I_0}$ is the only element that make Josephson junction circuits distinct from normal RLC circuits.

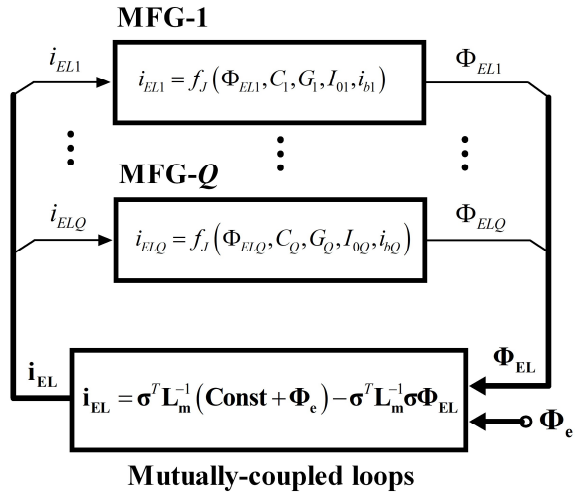

Fig. 14. General system model of MFG networks.

*2.6 Dynamics of MFG networks*

The general network equation in (13) is depicted with a system model shown in Fig. 14. This system model is easily simulated by using mathematic tools. This model exhibits the working principles of MFG networks as follows:
1) MFGs are flux chargers; they are supplied by both external bias currents and loop

currents, and generate flux outputs inside loops.

2) Loops are flux containers; they merge the flux outputs from MFGs and turn them into loop currents.

3) The external flux $\Phi_e$ applied to loops is turned into loop currents, and finally adjusts the total bias currents of MFGs.

*2.7 Magnetic-flux flow diagram*

To visualize the interactions between MFGs and loops, we invented an object-oriented diagram called magnetic-flux-flow diagram (MFF) [31]. The objects and their connections in MFF diagrams are illustrated in Fig. 15. a brick-shape block shown in Fig.15(a) represents an MFG; a circle shown in Fig. 15(b) is a loop; the symbol similar to the 'ground' in conventional circuit diagrams, represents the outer-loop [35], as shown in Fig. 15(c).

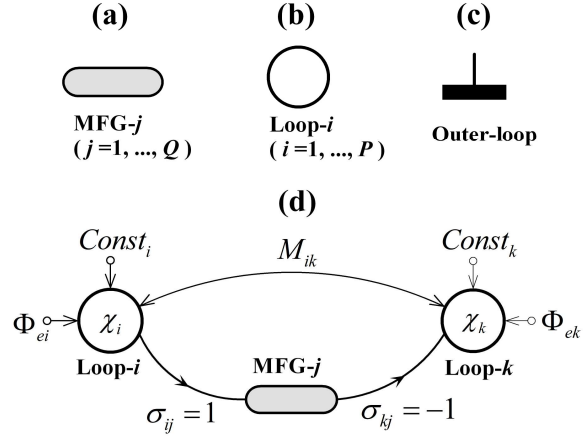

Fig. 15. Symbols of MFF diagrams and their connections: (a) abstract symbol of MFGs; (b) abstract symbol of loops; (c) symbol of outer-loop; (d) connections for loops and MFGs.

A simple MFF diagram is illustrated in Fig. 15(d), where MFGs and loops are interconnected according to the nonzero elements in **σ** and **L$_m$**. For example, a non-zero $\sigma_{ij} =1$ is turned into a directed line that connects Loop-$i$ with MFG-$j$, which arrow points to MFG-$j$; for $\sigma_{ik} = -1$, a directed line connect MFG-$j$ to Loop-$k$, with an arrow pointing to Loop-$k$; for a nonzero $M_{ik}$, a double-arrow line with two arrows pointing to loops is connected between Loop-$i$ and Loop-$k$. Moreover, nonzero *Const* and $\Phi_e$ are inputs of loops, and are represented as short lines with arrows entering loops.

The arrow indicates the polarity of the input applied to MFGs or loops; the input with an arrow entering MFGs or loops achieves a plus sign (+), otherwise, the input has a minus sign (–). Therefore, in Fig. 15(d), MFG-$j$ generates $-\Phi_{ELj}$ to Loop-$i$ and $+\Phi_{ELj}$ to Loop-$k$. Meanwhile, the arrow of directed lines depicts the direction of magnetic-flux flows between MFGs and loops.

Furthermore, we mark the value of $\chi$ inside the circle of loops to visualize the flux quanta transferring among loops. If MFG-$j$ achieves an increasement of one flux-quantum ($1\Phi_0$) in its flux output, this increasement will change $\chi_i$ to $\chi_i -1$ and the $\chi_k$ to $\chi_k + 1$, which looks like that MFG-$j$ transfers a flux-quantum from Loop-$i$ to Loop-$k$ along the directed lines.

MFF diagrams are the pictures of electric circuits viewed from magnetic fields, while the conventional diagrams are the pictures viewed from electric fields. MFF diagrams are complementary to the conventional circuit diagrams in the design and analysis of Josephson junction circuits.

*2.8 Comparison between MFF diagrams and conventional circuit diagrams*

Due to the charge-flux duality, a given electric circuit can be described with both the conventional circuit diagram and the MFF diagram, and can be analyzed using both the MNA method and the loop-current method accordingly. Two methods are mathematically equivalent in calculating the responses of circuit elements, but are different in depicting the overall behaviors of the circuit. The Conventional diagram and the MNA method depict the electric-charge flows inside the circuit; the MFF diagram and the flux-based loop-current method reveal the magnetic-flux flows inside the circuit.

A multiloop Josephson junction circuit shown in Fig. 16 (a) is used as an example for the comparison of two methods. The circuit consists of nine loops and twelve Josephson junctions; each loop, Loop-$j$ ($j$ =1, 2, …9), is applied with an external flux $\Phi_{ej}$ and trapped with flux quanta $n_j\Phi_0$. For the MNA method, the conventional circuit diagram is drawn as shown in Fig. 16(b), where the multiple flux inputs have to be internalized in branches as virtual flux sources, inductors, and transformers. However, when the phase inputs $n_j\Phi_0$ and $\Phi_{ej}$ for Loop-$j$ are treated as flux sources added to branches, there will be unexpected couplings between loops. For example, if $\Phi_{e5}$ is assigned to the branch between $\varphi_7$ and $\varphi_{11}$, it will also generate a current in Loop-6 (with a loop current $i_{m6}$); to cannel this effect, another $\Phi_{e5}$ have to be added to the branch between $\varphi_8$ and $\varphi_{12}$. Similarly, if the $n_5\Phi_0$ trapped by Loop-5 (with a loop current $i_{m5}$) is assigned to the branch between $\varphi_{10}$ and $\varphi_{11}$, another $n_5\Phi_0$ have to be inserted in the branch between $\varphi_{14}$ and $\varphi_{15}$ to cancel its effect on Loop-8 (with a loop current $i_{m8}$). Meanwhile, the conventional circuit diagram is inconvenience to describe all the mutual couplings between branches by using inductors and transformers.

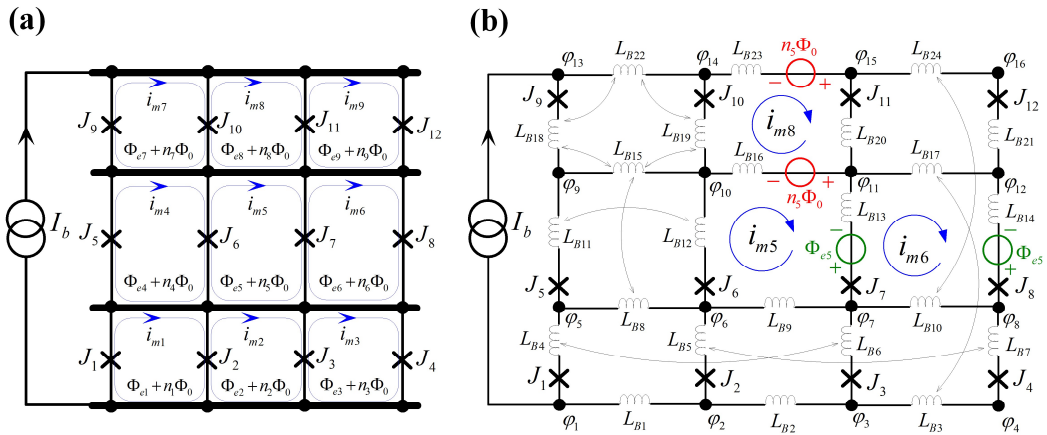

Fig. 16. (a) A multiloop Josephson junction array, and (b) the equivalent circuit emphasizing on the nodes and branches.

The conventional circuit diagram Fig. 16(b) looks clumsy in modeling the circuit with various flux inputs. During drawing the equivalent circuit, the electronics engineers must be familiar with the virtual nodal phases and the flux sources inserted in branches, and be aware of the FQL to make sure that those virtual components are properly placed in branches. Even though the FQL and Josephson effects are correctly described by the conventional circuit diagram, the circuit equations derived by the MNA method cannot describe how the flux quanta are distributed among loops for Josephson junction circuits, with nodal phases and branch currents.

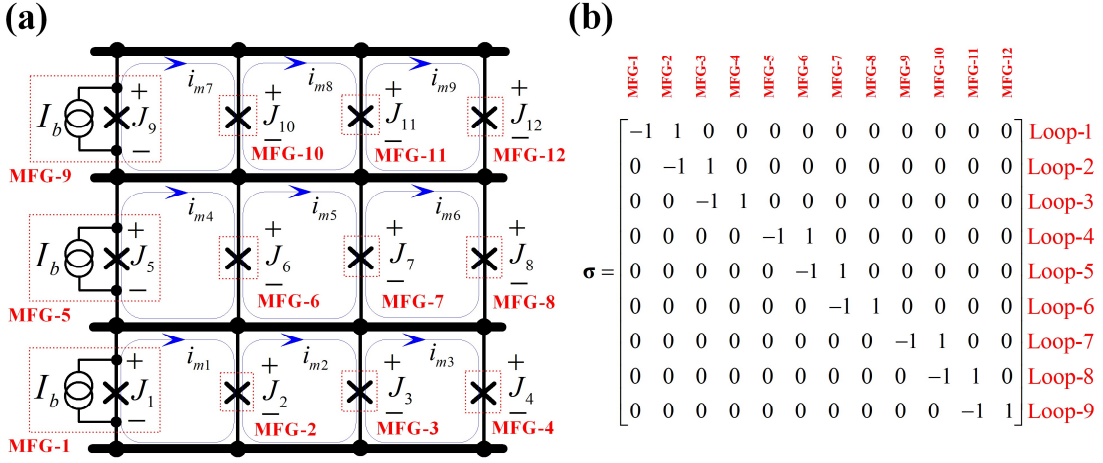

Fig. 17. (a) MFG network of the circuit shown in Fig. 16(a); (b) loop-to-MFG incidence matrix $\sigma$ extracted from the MFG network.

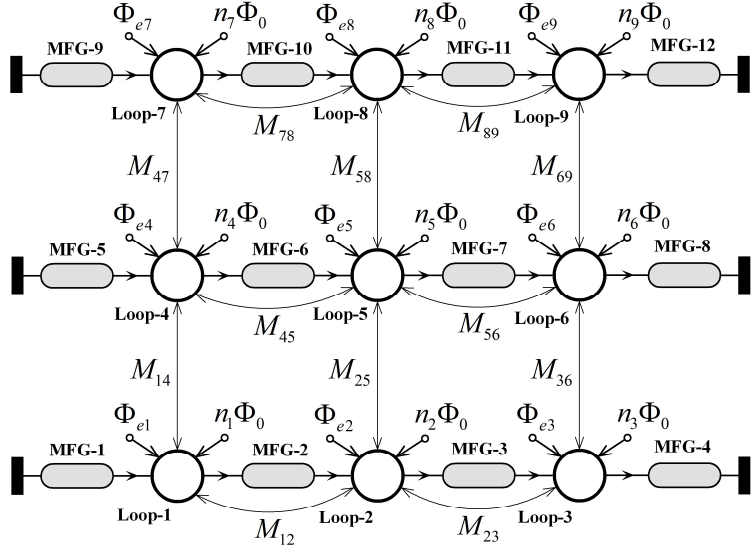

Fig. 18. MFF diagram of the circuit shown in Fig. 16(a).

The MFG network of the example is shown in Fig. 17(a), from which, we extract the matrix $\sigma$, as shown in Fig. 17(b), and extract the matrix $\mathbf{L_m}$ according to couplings between loops. With the elements of $\sigma$ and $\mathbf{L_m}$, we can simply set up the network equation and draw the MFF diagram for the Josephson junction circuit, as shown in Fig. 18, where three pipe lines transfer flux quanta from left to right, and loops are interconnected with mutual inductances. All those mutual inductances can be graphically represented with double-arrow lines between loops, and configured in the general network equation for numerical simulations.

Two methods have different performances in the analyses of Josephson junction circuits. Conventional circuit diagrams and the MNA method are not suitable for the Josephson junction circuits, because Josephson junction circuits work as magnetic-flux distribution networks more than electric-charge distribution networks. Our MFF diagrams and the flux-based loop-current method reveal the dynamics of magnetic-flux follows among loops and depict the image of circuits working as the magnetic-flux distribution networks, for both superconducting Josephson junction circuits and normal RLC circuits.

In summary, the conventional circuit diagram and the MFF diagram of an electric circuit provides two views of the circuit, like two sides of a coin. Conventional circuit diagrams depict the routes of electric-charge flows, while the MFF diagram depicts the routes of magnetic-flux flows inside the circuit. Their differences have been explained by the charge-flux duality principle shown in Fig. 5.

*2.9 Object-oriented circuit analysis methodology*

According to the charge-flux duality demonstrated in the reference [32], MFF diagrams can be renamed to develop electric-charge flow (ECF) or electric-flux flow (EFF) diagrams. The symbols and the connections for ECF diagrams are shown in Fig. 19, where electric-charge generators (ECGs) are dual to MFGs, and nodes are dual to loops. We can find that the ECF diagram is more concise than the conventional circuit diagram in depicting the electric-charge flows for an electric circuit, and the general network equation of ECG networks is more concise and intuitive than the circuit equations derived by the MNA method, in describing the dynamics of electric-charge distributions.

One symbology in MFF diagrams and ECF diagrams can be used to depict two pictures of electric circuits. This symbology can be powered by artificial intelligence to develop object-oriented circuit analysis methodologies in the future.

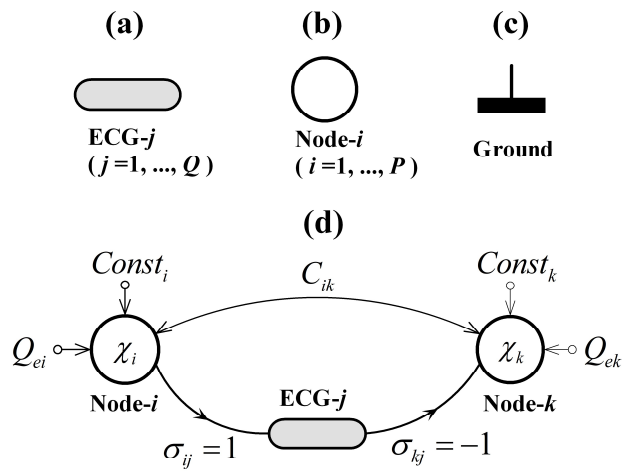

Fig. 19. Symbols of ECF diagrams and their connections: (a) abstract symbol of ECGs; (b) abstract symbol of nodes; (c) symbol of ground; (d) connections of ECGs and nodes.

# 3. SQUID Circuits Analyses

*3.1 Rf-SQUID*

The rf-SQUID circuit shown in Fig. 3(a) contains two MFGs, as shown in Fig. 20(a). The Josephson junction $J_1$ is defined as MFG-1, and the 50Ω impedance driven by the rf current source $i_{rf}(t)$ is defined as MFG-2. The MFG network is shown in Fig. 20(b); it consists of two loops, Loop-1 and Loop-2; two loops coupled with the mutual-inductance $M_{12}$. The system model of rf-SQUID is shown in Fig. 21. Two transfer functions of MFGs are associated through the linear flux-current functions of loops.

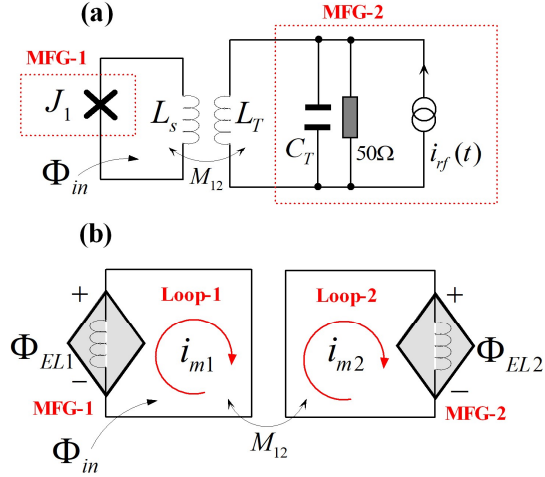

Fig. 20. (a) MFGs defined in rf-SQUID circuit, where $i_{b1} = 0$ and $i_{b2} = I_{rf}\sin(2\pi f_{rf}t)$, $I_{02} = 0$. (b) Equivalent MFG network, which has two MFGs contained in two loops.

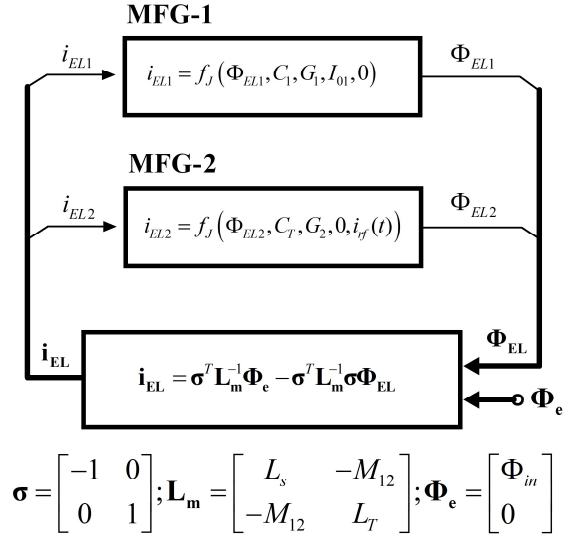

Fig. 21. System model of rf-SQUID circuit.

The MFF diagram of rf-SQUID is depicted in Fig. 22, where MFG-1 pumps fluxes into Loop-1 and MFG-2 drains fluxes out of Loop-2. Two loops are magnetically coupled through $M_{12}$. MFG-2 is biased with an ac current source, and generates a radio-frequency oscillating flux inside Loop-2; the oscillation of MFG-2 inside Loop-2 is interfered by the resonation of MFG-1 inside Loop-1, through $M_{12}$ between two loops. The external flux $\Phi_{in}$ adjusts the bias current of MFG-1, and alters the resonance of MFG-1 inside Loop-1, and thereby modulates the room-mean square (RMS) value of the voltage output of MFG-2. This is why we can obtain flux-modulated current-voltage characteristics from rf-SQUID.

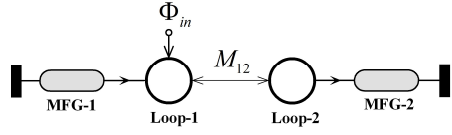

Fig. 22. MFF diagram of rf-SQUID circuit.

The current-voltage characteristics of rf-SQUID circuit are simulated, as shown in Fig. 23; they are similar to that measured from experiments [10]. In the simulation, all the circuit parameters are normalized with a reference resistor $R_0$ and a reference critical current $I_0$; for example, an inductance $L_x$ is normalized as $\beta_{Lx}$, $\beta_{Lx} = 2\pi I_0 L_x/\Phi_0$; a capacitance $C_x$ is normalized as $\beta_{Cx}$, $\beta_{Cx} = 2\pi I_0 R_0^2 C_x/\Phi_0$.

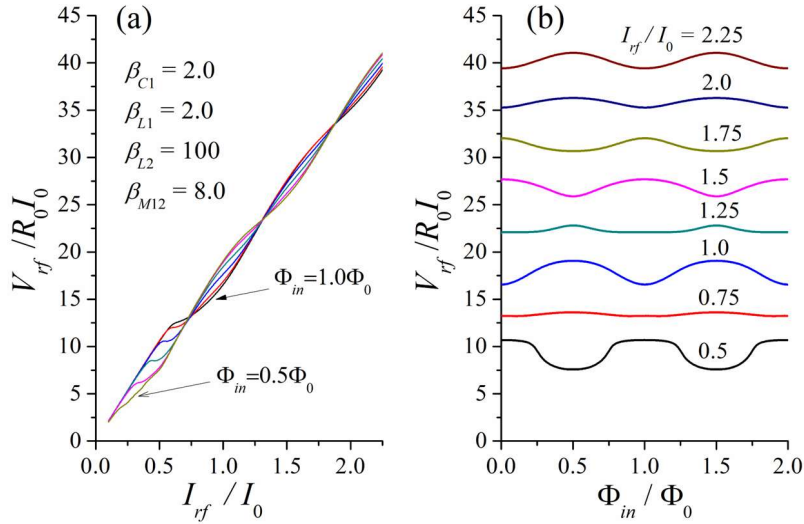

Fig. 23. Simulations of rf-SQUID circuit: (a) current-voltage curves and (b) flux-to-voltage curves, where $I_{01} = I_0$; $G_1 = 1/R_0$; $G_2 = 1/50$; $i_{rf} = I_{rf}\sin(2\pi f_{rf}t)$, $f_{rf} = 1/2\pi\sqrt{(L_1C_1)}$; $V_{rf}$ is the root-mean-square (RMS) value of $u_{EL2}$.

### 3.2 Dc-SQUID

The dc-SQUID circuit shown in Fig. 3(b) has two MFGs, each of which is biased with $I_b/2$, as shown in Fig. 24(a). The MFG network of dc-SQUID is shown in Fig. 24(b), where Loop-1 is the only loop that connects two MFGs.

The system model of dc-SQUID is shown in Fig. 25. MFG-1 is pulled by the loop-current of Loop-1 to keep pumping flux quanta into Loop-1, meanwhile, MFG-2 is pushed by Loop-1 to keep drawing flux quanta out of Loop-1. The average flow-rate of magnetic-flux flow through MFG-1 and MFG-2 is exactly the average voltage measured at two terminals of dc-SQUID.

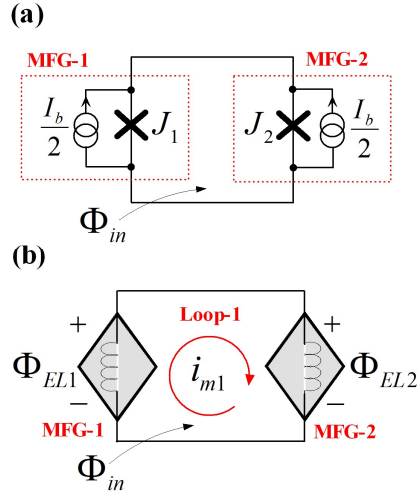

Fig. 24. (a) MFGs defined in dc-SQUID circuit, where $i_{b1} = i_{b2} = I_b/2$. (b) Equivalent MFG network, which has two MFGs embedded in one loop.

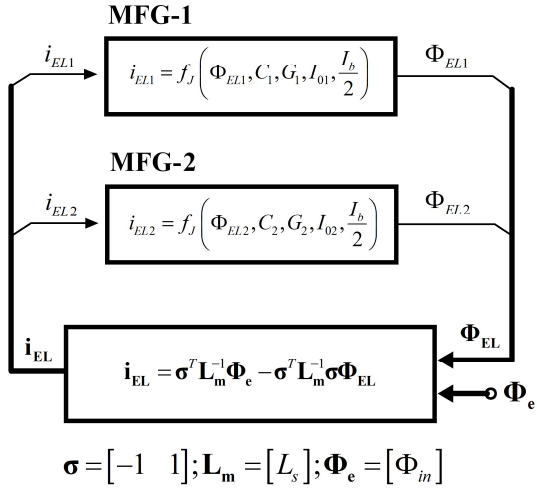

Fig. 25. System model of dc-SQUID.

The MFF diagram of dc-SQUID is shown in Fig. 26, where MFG-1 keep pumping flux quanta to Loop-1, and MFG-2 keep draining flux quanta out. The external flux $\Phi_{in}$ alters the bias currents of two MFGs, and finally modifies the current-voltage characteristic. This working principle is also verified in frequency domain [36].

The current-voltage characteristics and the flux-voltage curves of the dc-SQUID are shown in Fig. 27. They are same with the simulation results obtained by using phase-based analysis methods [37], [38].

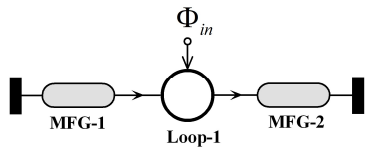

Fig. 26. MFF diagram of dc-SQUID.

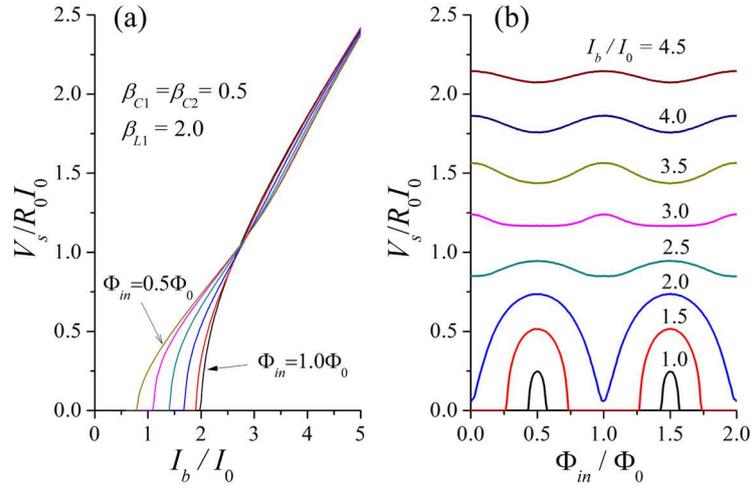

Fig. 27. Simulations of dc-SQUID: (a) current-voltage curves and (b) flux-to-voltage curves, where $I_{01} = I_{02} = I_0$; $G_1 = G_2 = 1/R_0$; $V_s$ is the average value of $u_{EL1}$ or $u_{EL2}$.

*3.3 Bi-SQUID*

The bi-SQUID circuit shown in Fig. 3(c) consists of three MFGs, as shown in Fig. 28(a). The MFG diagram has two loops that are directly coupled by sharing MFG-3, as shown in Fig. 28(b).

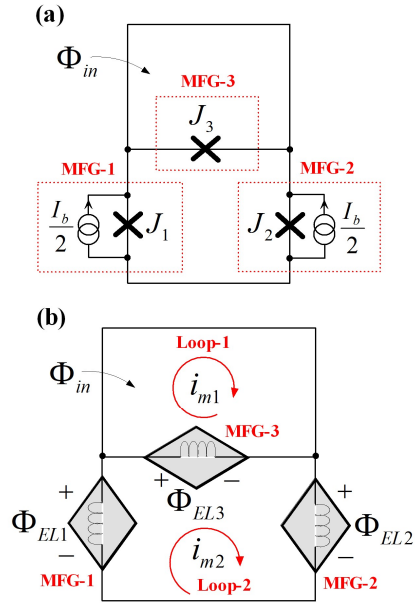

Fig. 28. (a) MFGs defined in bi-SQUID, where $i_{b1} = i_{b2} = I_b/2$ and $i_{b3} = 0$. (b) Equivalent MFG network, which has three MFGs contained by two loops.

The system model of bi-SQUID is shown in Fig. 29. The corresponding MFF diagram is depicted in Fig. 30; the bi-SQUID is constituted by connecting a rf-SQUID to a dc-SQUID; the rf-SQUID consisting of Loop-1 and MFG-3 works as a snubber that absorbs the flux in loop-2 and modifies the flow-rate of MFG-1. The external flux $\Phi_{in}$ modifies the snubber effect of MFG-3 on the Loop-2 that results in the linearized flux-to-voltage characteristics, as illustrated in Fig. 31. The simulations are agreed well with the typical experimental results in reference [13].

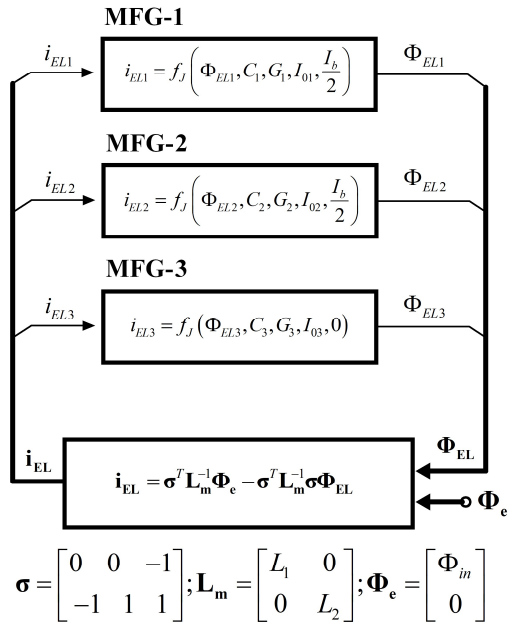

Fig. 29. System model of bi-SQUID.

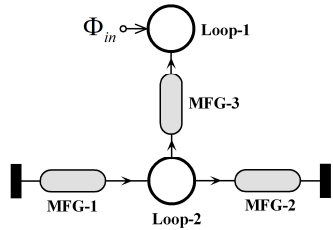

Fig. 30. MFF diagram of bi-SQUID.

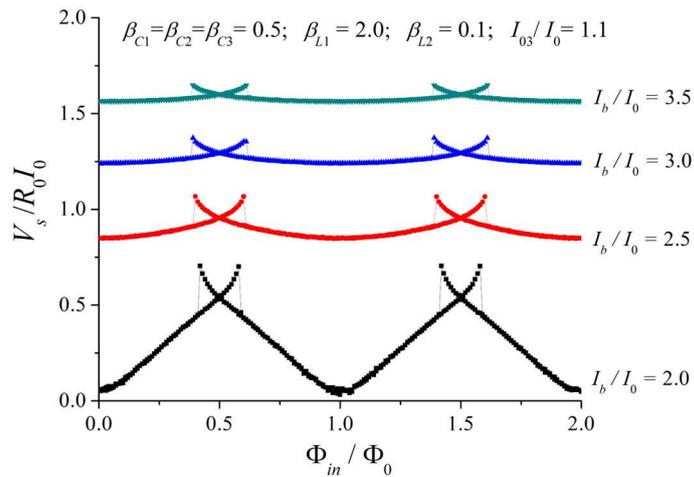

Fig. 31. Simulations of flux-to-voltage characteristics for bi-SQUID, where, $I_{01} = I_{02} = I_0$; $I_{03} = 1.1 I_0$; $G_1 = G_2 = G_3 = 1/R_0$; $\beta_{M12} = 0$; $V_s$ is the average value of $u_{EL1}$ or $u_{EL2}$.

## 3.4 Dro-SQUID

The dro-SQUID circuit shown in Fig. 3(d) is redrawn as shown in Fig. 32(a), where the shunt resistor $R_a$ accompanied with current source $I_b$ is defined as MFG-4. The MFG diagram of dro-SQUID is shown in Fig. 32(b), where there are two loops electrically coupled through sharing the branch of MFG-2.

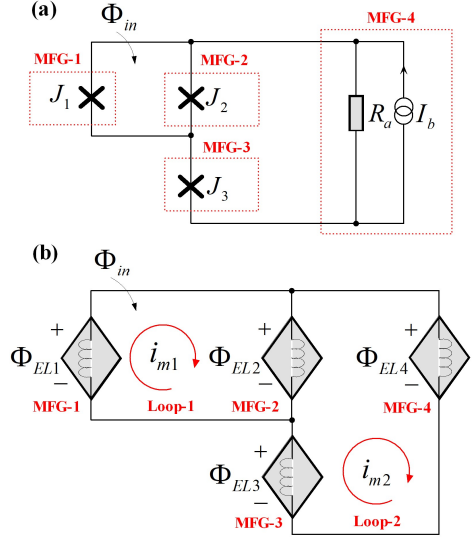

Fig. 32. (a) MFGs defined dro-SQUID, where $i_{b1}= i_{b2} =i_{b3} = 0$; in MFG-4, $i_{b4} = I_b$, $I_{04} = 0$ and $C_4 = 0$.
(b) MFG network of dro-SQUID, which has four MFGs contained by two loops.

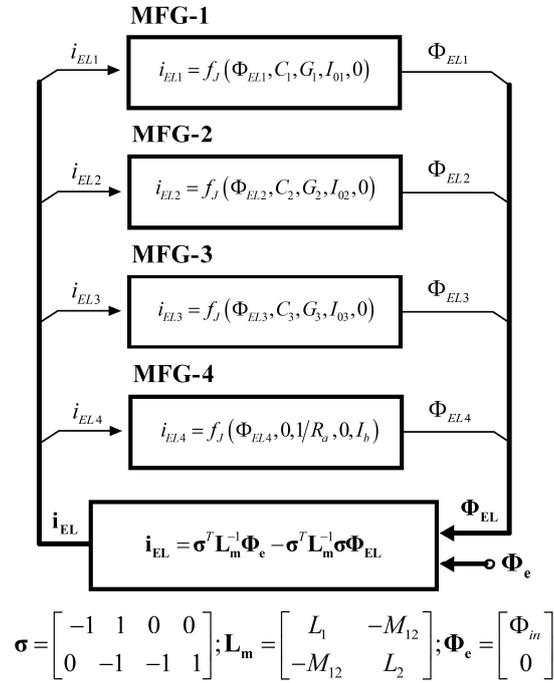

Fig. 33. System model of dro-SQUID.

The system model of dro-SQUID is shown in Fig. 33. The corresponding MFF diagram is shown in Fig. 34, where MFG-4 works as a drain to Loop-2, while MFG-2 and MFG-3 are two sources. The flux-flow is supplied by either MFG-3 or MFG-2; if it flows from MFG-3, the voltage measured at MFG-3 will be nonzero; if it flows through MFG-2, the voltage output of MFG-3 is zero. The external flux $\Phi_{in}$ applied to Loop-1 alters the bias current of MFG-2, and switches the flux-flow through MFG-3. Consequently, the output of dro-SQUID at two terminals of MFG-3 will achieve square-wave-shaped flux-voltage characteristics [39], as exhibited in Fig. 35.

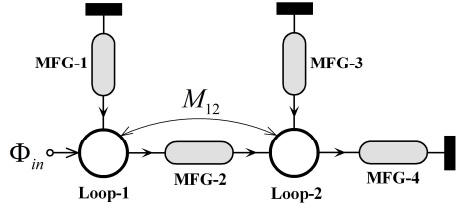

Fig. 34. MFF diagram of dro-SQUID.

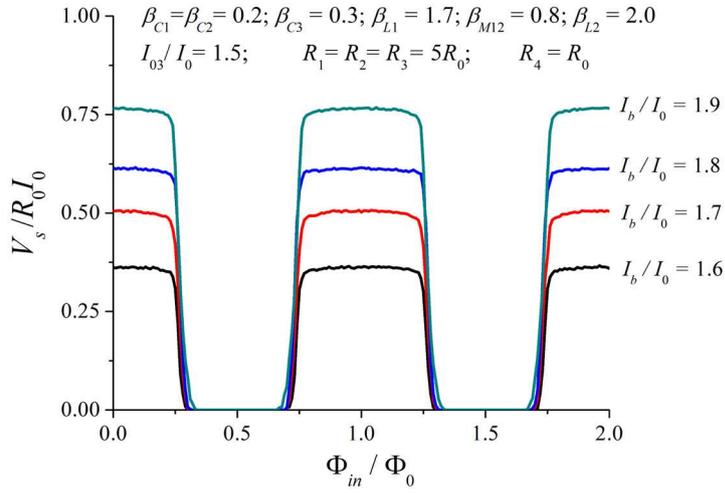

Fig. 35. Simulations of flux-to-voltage characteristics for dro-SQUID, where $I_{01} = I_{02} = I_0$; $I_{03} = 1.5I_0$; $G_1 = G_2 = G_3 = 1/(5R_0)$; $G_4 = 1/R_0$; $V_s$ is the average value of $u_{EL3}$.

*3.5 SQIF*

The SQIF shown in Fig. 1(e) is redrawn with MFG model, as shown in Fig. 36(a); its MFF diagram is shown in Fig. 36(b). The system model of SQIF is shown in Fig. 37, where MFG-$N$ is biased by $I_b$. The MFF diagram is shown in Fig. 38; it looks like a pipeline, where MFG-$N$ drains the fluxes out of Loops connected in series. The external flux inputs $\Phi_{in1}$ to $\Phi_{in(N-1)}$ alter the bias currents of MFGs that will consequently modulate the average flow-rate of the pipeline.

*3.6 SQUID Array*

The SQUID-array shown in Fig. 1(f) is redrawn as shown in Fig. 39(a), where each Josephson junction is defined as an MFG, and the bias current accompanied with a load resistor $R_L$ is defined as MFG-$(N+1)$. The MFF diagram intuitively depicts the magnetic-flux flow inside the SQUID-array, as shown in Fig. 39(b); magnetic-fluxes stored in Loop-1 – Loop-$P$ are merged in Loop-$(P+1)$, and are finally absorbed by MFG-$(N+1)$. The external flux inputs, namely $\Phi_{in1} \sim \Phi_{inP}$, change the bias current of

MFGs, and thus modulate the flow-rate through the final MFG. The dynamics of SQUID-array is exhibited by the system model shown in Fig. 40.

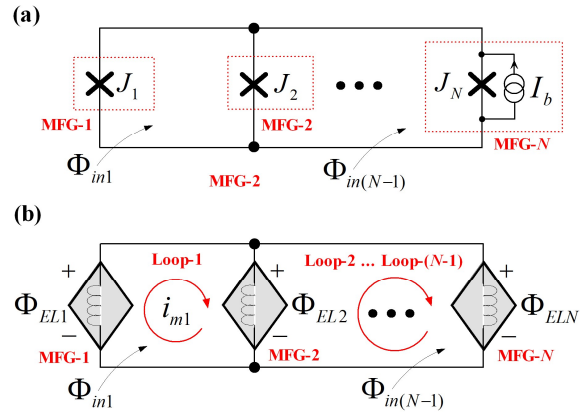

Fig. 36. (a) MFGs defined in SQIF circuit; (b) MFG network of SQIF.

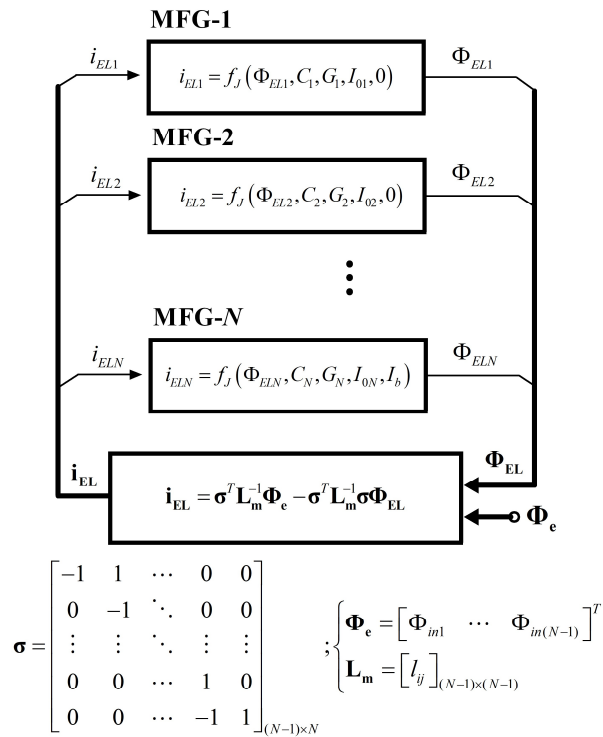

Fig. 37. System model of SQIF.

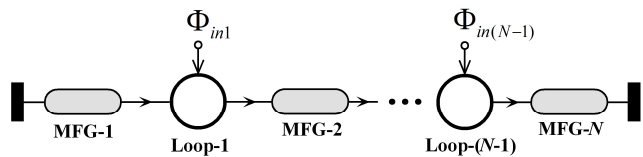

Fig. 38. MFF diagram of SQIF

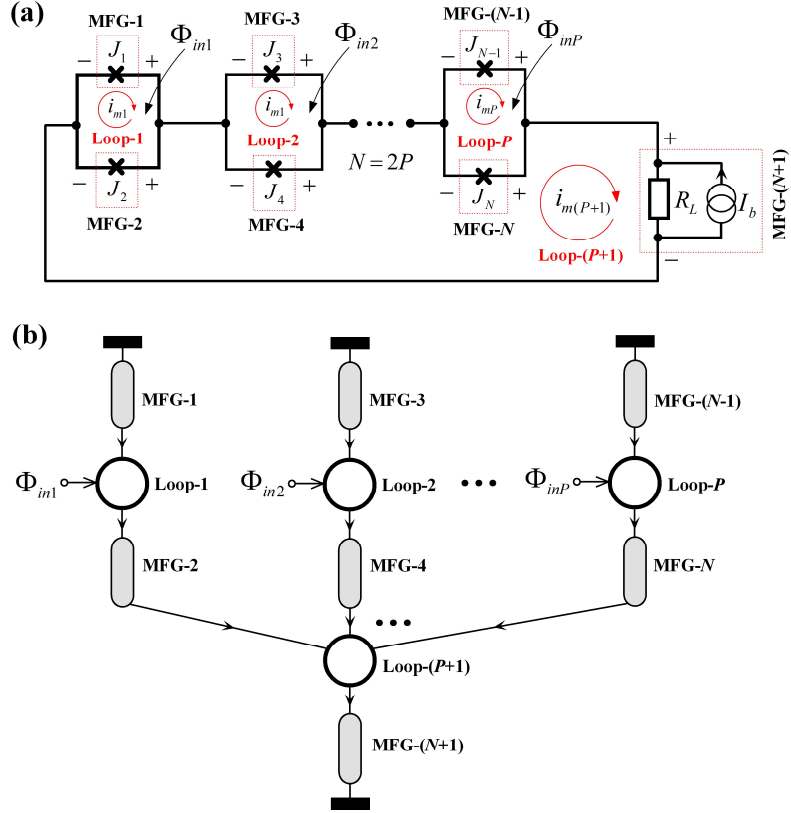

Fig. 39. (a) MFGs and loops defined in SQUID-array; (b) MFF diagram of SQUID-array.

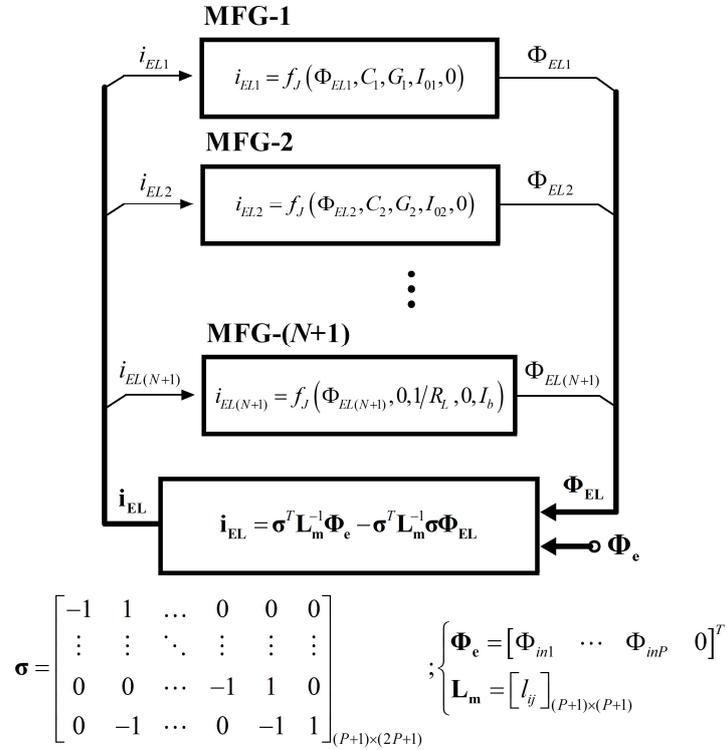

Fig. 40. System model of SQUID-array.

# 4. SFQ Circuits Analyses

*4.1 TTL-to-SFQ convertor*

The TTL-to-SFQ convertor shown in Fig. 2(a) is redrawn with the MFG model, as shown in Fig. 41(a). It contains two MFGs and one loop; the external flux $\Phi_{in}$ applied TTL logics is the input to Loop-1; the flux output of MFG-2 is the output for subsequent circuits. The MFF diagram of this TTL-to-SFQ convertor is shown in Fig. 41(b), where MFG-1 is used to generate a flux-quantum in Loop-1, and MFG-2 is used to pump the flux-quantum to next stages; the $\Phi_{in}$ controls the trigger of MFGs.

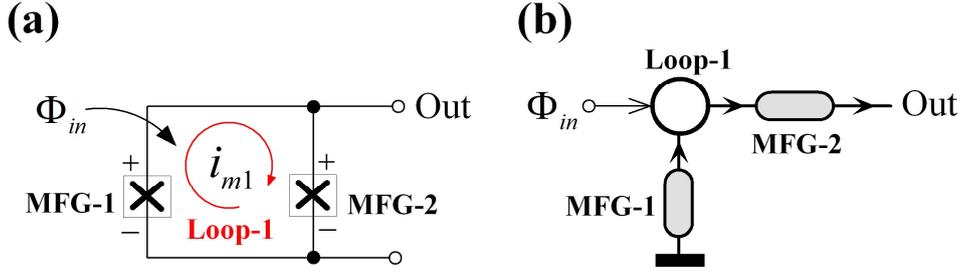

Fig. 41. (a) TTL-to-SFQ convertor consisting of two MFGs and one loop; (b) MFF diagram of TTL-to-SFQ convertor.

The operation of TTL-to-SFQ convertor is illustrated in Fig. 42, where one flux-quantum contained in loops is represented with a red ball. The MFF diagram clearly depict how the TTL-to-SFQ convertor transform a TLL pulse into a flux-quantum contained in the loop:

1) At first, two MFGs stay still when $\Phi_{in} = 0$, since the bias currents of MFG-1 and MFG-2 are less than their critical currents, as shown in Fig. 42(a).

2) When $\Phi_{in}$ is altered from 0 to $-\Phi_0$, the loop current in Loop-1 decreases and triggers MFG-1 to generate a flux-quantum in Loop-1, as shown in Fig. 42(b).

3) When $\Phi_{in}$ returns to 0, the loop-current of Loop-1 will trigger MFG-2 to transfer the flux-quantum from Loop-1 to the subsequent loop, as shown in Fig. 42(c).

It is shown that, in SFQ circuits, loops are the entities of SFQ logic bits, and their '0' and '1' states are defined by the value of $\chi$.

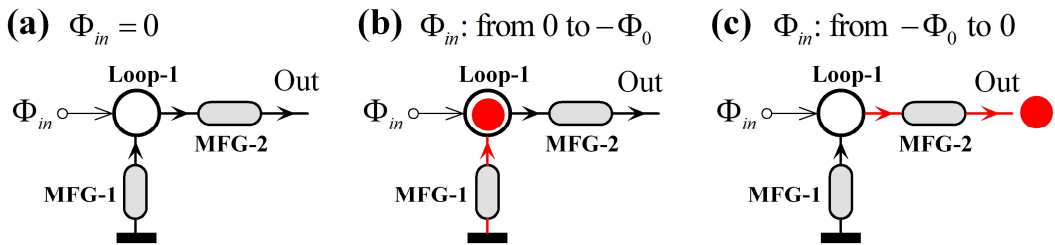

Fig. 42. Operations of TTL-to-SFQ convertor, where $\Phi_{in}$ is generated by a TTL input: (a) Loop-1 is empty when $\Phi_{in}$ is zero for the low-level of the TTL input; (b) a flux-quantum is induced by MFG-1 to Loop-1, when $\Phi_{in}$ is turned to $-\Phi_0$ by the high-level of the TTL input; (b) the flux-quantum in Loop-1 is pumped out by MFG-2, when $\Phi_{in}$ returns to zero with the TTL input back to low-level.

*4.2 Josephson Transmission Line*

The JTLs are used to transfer the flux quanta generated by TTL-to-SFQ convertors to SFQ logic gates. The JTL unit shown in Fig. 2(b) is turned into a MFG network, as shown in Fig.43(a); its MFF diagram looks like a pipe that transfers flux quanta from loops to loops, as shown Fig. 43(b). The working principle of the JTL is vividly depicted in Fig. 44, where MFGs relay a flux-quantum from one loop to another.

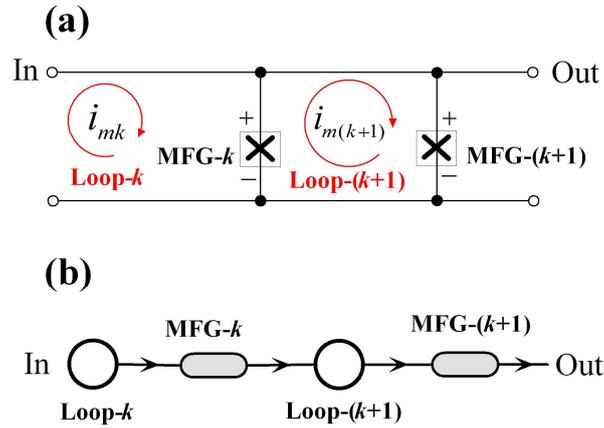

Fig. 43. (a) MFGs and loops defined in JTL; (b) MFF diagram of JTL.

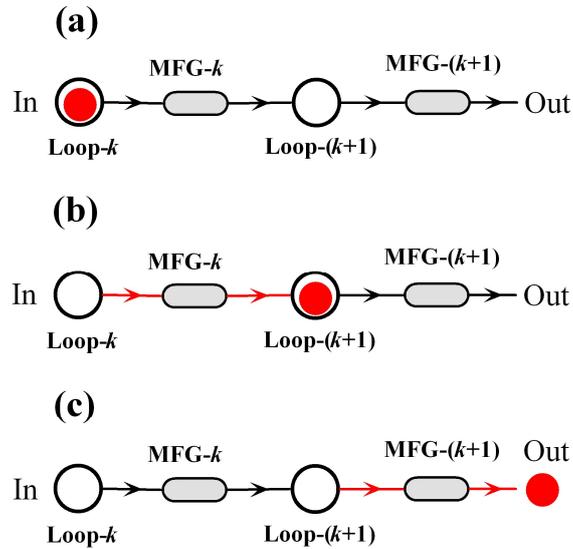

Fig. 44. Operations of JTL: (a) a flux-quantum is pumped in Loop-$k$; (b) it is transferred by MFG-$k$ to Loop-($k$+1), and (c) is transmitted out by MFG-($k$+1), continuously.

*4.3 Josephson D-type Flip-Flop*

For the DFF circuit shown in Fig. 2(c), its MFG network is shown in Fig. 45(a). There are three loops containing four MFGs. The interactions between loops and MFGs are exhibited by the MFF diagram shown in Fig. 45(b).

The principle of the Josephson DFF is illustrated in Fig. 46. At first, a flux-quantum is pumped into Loop-2 by the MFG from former stage, as shown in Fig. 46(a); it is transferred to Loop-3 by MFG-3, as shown in Fig. 46(b), and will be stored in Loop-3, as long as Loop-1 is empty. When Loop-1 receives a flux-quantum generated by the clock signal, as shown in Fig. 46(c), the flux quanta in Loop-1 and Loop-3 will be merged into one flux-quantum and pumped out by MFG-4, as shown in Fig. 46(d). If Loop-3 is empty, the flux-quantum pumped into Loop-1 will not trigger the MFG-4, and will be dumped by MFG-1; in this case, there will be no flux-quantum output at MFG-4. The MFF diagram shows that the Josephson DFF synchronizes the input with the clock signals, similar with the semiconductor DFF logics.

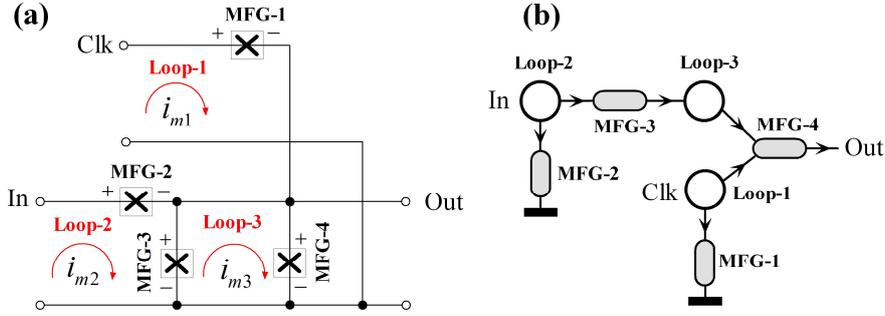

Fig. 45. (a) MFGs and Loops in Josephson DFF, and (b) the MFF diagram.

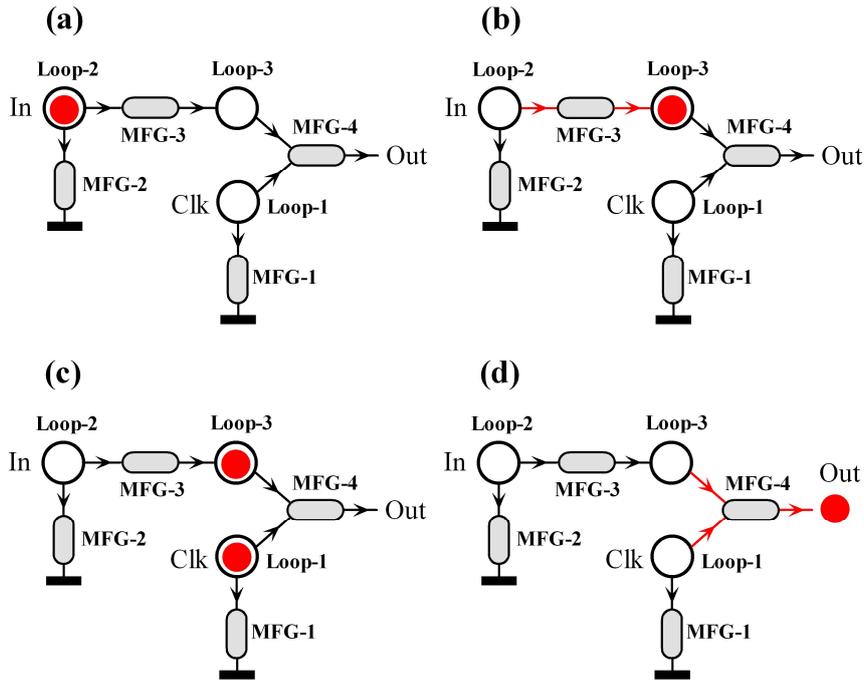

Fig. 46. Dynamics of Josephson DFF: (a) a flux-quantum is pumped in Loop-2; (b) it is transferred into Loop-3; (c) another flux-quantum is pumped in Loop-1 by clock signals; (d) flux quanta in Loop-3 and Loop-4 are merged into one flux-quantum and pumped out by MFG-4. More details can be found in reference [31].

*4.4 Josephson AND-gate*

The Josephson AND-gate shown in Fig. 2(d) is redrawn as the MFG network shown in Fig. 47(a), in which, MFG-1 is embedded in Loop-1, while MFG-2 is in Loop-2; MFG-3 are shared by Loop-1 and Loop-2.

The MFF diagram shown in Fig. 47(b) vividly depicts the AND logic. If Loop-1 receives a flux-quantum when Loop-2 is empty, as shown in Fig. 48(a), the flux-quantum in Loop-1 will be absorbed by MFG-1, and there will no output at MFG-3, as shown in Fig. 48(b). When two flux quanta shown in Fig. 48(c) are pumped into Loop-1 and Loop-2 at the same time, they will be merged into one flux-quantum and pumped out by MFG-3, as illustrated in Fig. 48(d). Therefore, MFG-3 will pump a flux-quantum out, only when two loops are in '1' state at the same time; this is exactly the logic of AND circuits.

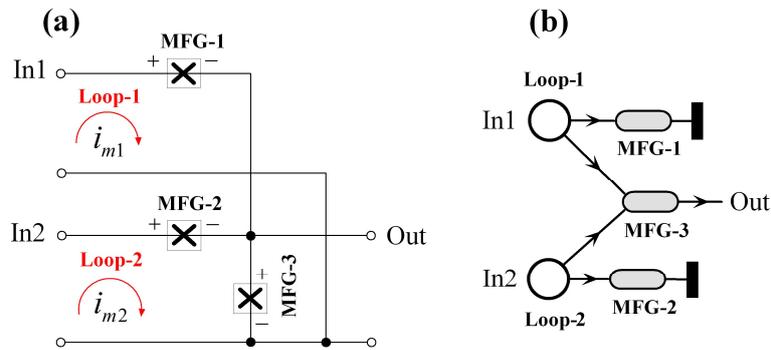

Fig. 47. (a) MFGs and loops in Josephson AND-gate; (b) MFF diagram of Josephson AND-gate.

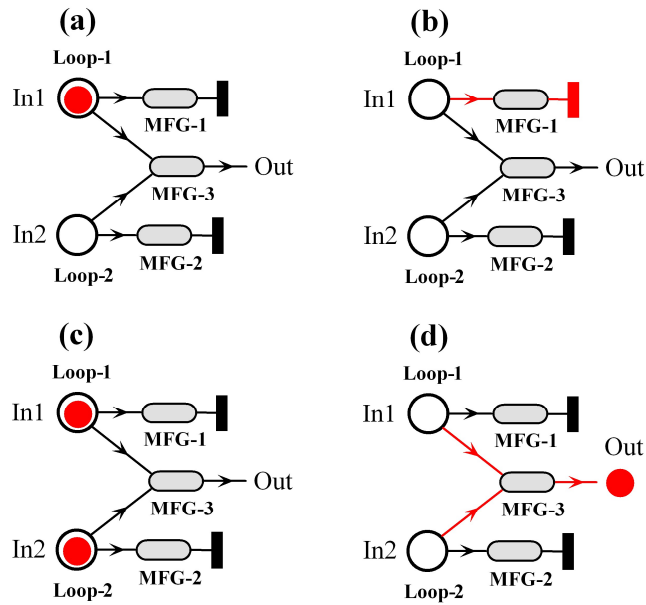

Fig. 48. Operations of Josephson AND-gate. (a) if a flux-quantum flows into Loop-1 while Loop-2 is empty, (b) it will be dumped by MFG-1; (c) if Loop-1 and Loop-2 receive a flux-quantum at the same time, (d) two flux quanta will be merged and pumped out by MFG-3. More details can be found in reference [31].

*4.5 Josephson OR-gate*

The Josephson OR-gate shown in Fig. 2(e) is modeled as the MFG network shown in Fig. 49(a). Its MFF diagram is depicted in Fig. 49(b). Loop-1 and Loop-2 are two input loops, their flux quanta are pumped out by MFG-4 according to the OR logic.

The principle of Josephson OR-gate is illustrated in Fig. 50. When a flux-quantum is pumped in Loop-1, it will decrease the loop-current of Loop-2 through $M_{12}$, as shown in Fig. 50(a). The decreased loop-current of Loop-2 will trigger MFG-2 to induce a flux-quantum to Loop-2, as shown in Fig. 50(b); when both loops are in '1' state, two flux quanta will be merged into one and pumped out by MFG-4, as shown in Fig. 50(c); they will also be dumped by MFG-3, if MFG-4 is blocked by the subsequent loop. Therefore, the flux-flow depicted by this MFF diagram implements exactly the OR logic which output is '1' when either input is '1'.

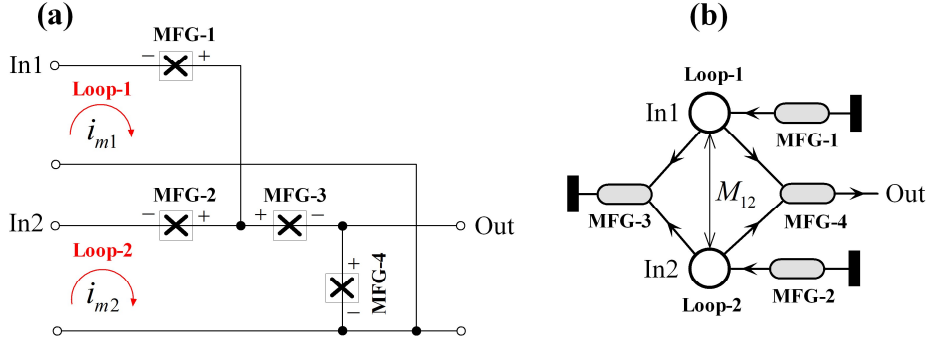

Fig. 49. (a) MFGs and loops in Josephson OR-gate; (b) MFF diagram of Josephson OR-gate.

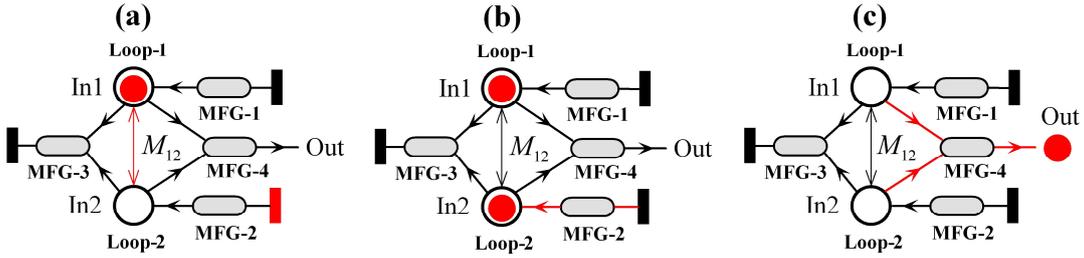

Fig. 50. Principles of the Josephson OR-gate: (a) a flux-quantum is pumped in Loop-1, while Loop-2 is empty; (b) it lowers the loop-current of Loop-2 through $M_{12}$ to trigger MFG-2 to generate a flux-quantum in Loop-2; (c) two flux quanta in Loop-1 and Loop2 will be merged into one flux-quantum and pumped out by MFG-4. More details can be found in reference [31].

*4.6 Josephson NOT-gate*

The Josephson NOT-gate shown in Fig. 2(f) consists of four loops and five MFGs, as shown in Fig. 51(a). The interactions between loops and MFGs are exhibited by the MFF diagram shown in Fig. 51(b). The MFF diagram is symmetric except that there is a negative mutual-inductance $M_{23}$ between Loop-2 and Loop-3. The flux-quantum in Loop-1 will be drawn out by either MFG-1 or MFG-2, being switched by the flux-quantum in Loop-2.

The working principles of the Josephson NOT-gate are demonstrated in Fig. 52. Depending on the negative $M_{23}$, the flux-quantum in Loop-2 will lower the loop-current in Loop-3 and raise the current flowing into MFG-1 that make MFG-1 be more easily triggered than MFG-2, as shown in Fig. 52(a). When a flux-quantum is pumped in Loop-1 by clock signals, it will firstly trigger MFG-1 and flow to Loop-3, as shown in Fig. 52(b). The flux-quantum accompanied with the flux-quantum in Loop-2 is finally absorbed by MFG-3, and there will no outputs at MFG-2, as shown in Fig. 52(c). If Loop-2 is empty, its flux coupled to Loop-3 through $M_{23}$ disappears; the loop-current in Loop-3 remains zero, and MFG-1 will be less easily triggered than MFG-2, as shown in Fig. 52(d). When Loop-1 receives a flux-quantum, MFG-2 will be firstly triggered to transfer the flux-quantum from Loop-1 to Loop-4, and pump out one flux-quantum at the same time, as shown in Fig. 52(e). The flux-quantum in Loop-4 will be finally absorbed by MFG-4, during which, the flux-quantum in Loop-2 drained by MFG-4 is compensated by MFG-5.

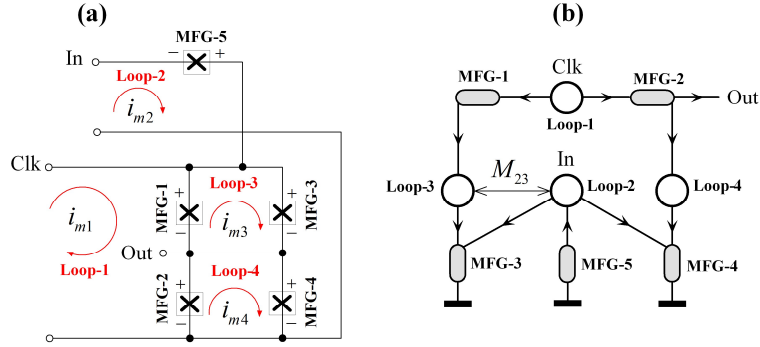

Fig. 51. (a) MFGs and loops defined in a Josephson NOT-gate; (b) MFF diagram of the Josephson NOT-gate.

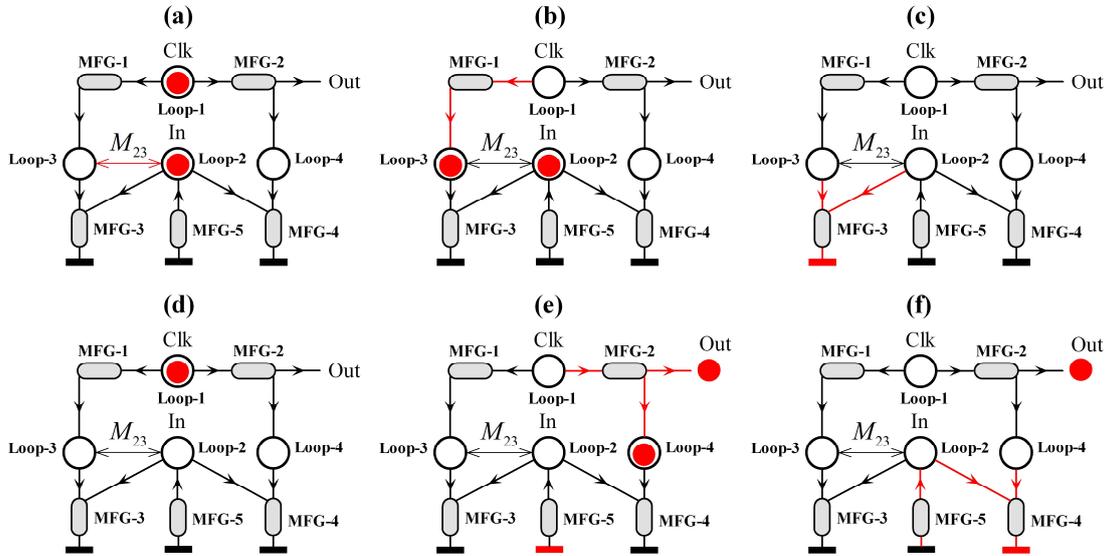

Fig. 52. Principles of Josephson NOT-gate, where Loop-1 is the input of clock signals and Loop-2 is the input that controls the output. (a) a flux-quantum input is already in Loop-2 before Loop-1 receives a flux-quantum, it will lower the loop-current of Loop-3 through $M_{23}$; (b) the flux-quantum in Loop1 is transferred by MFG-1 to Loop-3; (c) two flux quanta in Loop-2 and Loop-3 are dumped by MFG-3, and there will be no flux-quantum pumped out from the NOT-gate; (d) Loop-1 receives a flux-quantum, and Loop-2 is empty; (e) the flux-quantum in Loop-1 is duplicated by MFG-2; one is transferred to Loop-4, another one is pumped out as the output of the NOT-gate; (f) the flux-quantum in Loop-4 is finally absorbed by MFG-4, accompanied with a flux-quantum from Loop-2 which is generated by MFG-5. More details can be found in reference [31].

The NOT-logic depicted by this MFF diagram is that the Josephson NOT-gate will *not* output any clock signal until the input loop is in '0' state. Without the illustration of the MFF diagram shown in Fig. 52, it will be difficult to read the NOT-logic directly from the conventional circuit diagram shown in Fig. 1(f).

## 5. Discussion and Conclusion

We present a flux-based circuit theory to unify superconducting Josephson junction circuits and normal RLC circuits. This theory includes:

1) an MFG model used to unify Josephson junctions and normal RC components;

2) an MFG network applied to abstract both Josephson junction circuits and normal RLC circuits;

3) a general network equation derived to describe the dynamics of both superconducting and normal MFG networks;

4) a symbology of MFF diagram invented to graphically depict the working principles of magnetic-flux flows for Josephson junction circuits.

The applications of our flux-based circuit theory in the analyses of SQUIDs and SFQ circuits demonstrate that, Josephson junction circuits are magnetic-flux distribution networks where flux-quantum flows are rectified by Josephson junctions; their working principle are better to be depicted with MFF diagrams rather than conventional circuit diagrams. Our MFF diagrams are complementary to the conventional diagrams in the design and analyses of Josephson junction circuits, since the conventional circuit diagrams are supported by the electronic-design-automation (EDA) tools for layout and physical implementations. Moreover, the symbology of our MFF diagrams enables the objected-oriented circuit analysis methodology that may be applied by the circuit design tools powered by artificial intelligence in the future.

## Acknowledgements